\documentclass[a4paper,12pt]{article}
\usepackage{graphicx, rotating}
\usepackage{ifpdf}
\ifpdf
\usepackage{hyperref, pdfsync, epstopdf}    % This is for pdftex
\else
\usepackage[dvips,bookmarks]{hyperref}  % This is for arXiv.org
\fi
\hypersetup{colorlinks,bookmarksopen,bookmarksnumbered,citecolor=verdes,
linkcolor=blus,pdfstartview=FitH,urlcolor=rossos}
\def\hhref#1{\href{http://arxiv.org/abs/#1}{#1}} % in bibliography
\usepackage{amsmath}
\usepackage{slashed}

\renewcommand{\theequation}{\thesection.\arabic{equation}}

 \newcommand{\gr}{\ensuremath{\Psi}}
\newcommand{\gl}{\ensuremath{\lambda}}
\newcommand{\sq}{\ensuremath{\tilde{q}}}

\newcommand{\beq}{\begin{equation}}
\newcommand{\eeq}{\end{equation}}
\newcommand{\fig}[1]{~\ref{fig:#1}}

\newcommand{\Op}{{\cal O}}

\newcount\Mac  \Mac=1  % devo mettere Mac=1 se sto lavorando sul file Mac
\newcommand{\ifMac}[2]{\ifnum\Mac=1 #1 \else #2 \fi}
\def\putps(#1,#2)(#3,#4)#5#6{\ifnum\Mac=1 \put(#1,#2){\special{picture #5}}
\else  \put(#3,#4){\includegraphics{#6}} \fi}

\newcommand{\riga}[1]{\noalign{\hbox{\parbox{\textwidth}{#1}}}\nonumber}

\newcommand{\One}{\hbox{1\kern-.24em I}}

\renewcommand{\Im}{\mathop{\rm Im}}

\newcommand{\GeV}{\,{\rm GeV}}
\newcommand{\TeV}{\,{\rm TeV}}

\newcommand{\eV}{\,{\rm eV}}
\newcommand{\NP}{Nucl. Phys.}
\newcommand{\PRL}{Phys. Rev. Lett.}
\newcommand{\PL}{Phys. Lett.}
\newcommand{\PR}{Phys. Rev.}
\newcommand{\eq}[1]{~{\rm (\ref{eq:#1})}}

\newcommand{\Pl}{P\hspace{-1.5ex}/\,}

\newcommand{\ds}{\partial\!\!\!\raisebox{2pt}[0pt][0pt]{$\scriptstyle/$}\,}

\newcommand{\Ksl}{K\hspace{-1.6ex}/}
\newcommand{\Ssl}{S\hspace{-1.4ex}/\,}

\newcommand{\lascia}[1]{}
\makeatletter
\def\art{\@ifnextchar[{\eart}{\oart}}
\def\eart[#1]#2#3#4#5#6{{\rm #2}, {#3 #4} {\rm (#6) #5} [{\hhref{#1}}]}
\def\hepart[#1]#2{{\rm #2, \hhref{#1}}}
\newcommand{\oart}[5]{{\rm #1}, {#2 #3} {\rm (#5) #4}}

\newcounter{alphaequation}[equation]
\def\thealphaequation{\theequation\hbox to
0.6em{\hfil\alph{alphaequation}\hfil}}
\def\eqnsystem#1{
\def\@eqnnum{{\rm (\thealphaequation)}}
\def\@@eqncr{\let\@tempa\relax \ifcase\@eqcnt \def\@tempa{& & &} \or
  \def\@tempa{& &}\or \def\@tempa{&}\fi\@tempa
  \if@eqnsw\@eqnnum\refstepcounter{alphaequation}\fi
\global\@eqnswtrue\global\@eqcnt=0\cr}
\refstepcounter{equation} \let\@currentlabel\theequation \def\@tempb{#1}
\ifx\@tempb\empty\else\label{#1}\fi
\refstepcounter{alphaequation}
\let\@currentlabel\thealphaequation
\global\@eqnswtrue\global\@eqcnt=0 \tabskip\@centering\let\\=\@eqncr
$$\halign to \displaywidth\bgroup \@eqnsel\hskip\@centering
$\displaystyle\tabskip\z@{##}$&\global\@eqcnt\@ne
\hskip2\arraycolsep\hfil${##}$\hfil& \global\@eqcnt\tw@\hskip2\arraycolsep
$\displaystyle\tabskip\z@{##}$\hfil
\tabskip\@centering&\llap{##}\tabskip\z@\cr}

\def\endeqnsystem{\@@eqncr\egroup$$\global\@ignoretrue} \makeatother

\newcommand{\mb}[1]{\mbox{\normalsize\boldmath $#1$}}

\def\Lag{{\cal L}}
\def\SU{{\rm SU}}

\def\Tr{\mathop{\rm Tr}}
\def\circa#1{\,\raise.3ex\hbox{$#1$\kern-.75em\lower1ex\hbox{$\sim$}}\,}

\usepackage{multicol}
\usepackage{color}
\definecolor{rosso}{cmyk}{0,1,1,0.4}
\definecolor{rossos}{cmyk}{0,1,1,0.55}
\definecolor{rossoc}{cmyk}{0,1,1,0.2}
\definecolor{blu}{cmyk}{1,1,0,0.3}
\definecolor{blus}{cmyk}{1,1,0,0.6}
\definecolor{bluc}{cmyk}{1,1,0,0.1}
\definecolor{verde}{cmyk}{0.92,0,0.59,0.25}
\definecolor{verdec}{cmyk}{0.92,0,0.59,0.15}
\definecolor{verdes}{cmyk}{0.92,0,0.59,0.4}
\definecolor{grigio}{cmyk}{0,0,0,0.07}
\definecolor{rosa}{cmyk}{0,0.1,0.1,0.02}
\definecolor{rosino}{cmyk}{0,0.05,0.05,0.02}
\definecolor{rosas}{cmyk}{0,0.3,0.25,0.05}
\definecolor{celeste}{cmyk}{0.1,0,0,0.02}
\definecolor{giallino}{cmyk}{0,0,0.4,0.02}
\definecolor{rosso}{cmyk}{0,1,1,0.4}
\definecolor{rossos}{cmyk}{0,1,1,0.55}
\definecolor{rossoc}{cmyk}{0,1,1,0.2}
\definecolor{blu}{cmyk}{1,1,0,0.3}
\definecolor{bluc}{cmyk}{1,1,0,0.1}
\definecolor{blucc}{cmyk}{0.7,0.5,0,0}
\definecolor{viola}{cmyk}{0,1,0,0.6}
\definecolor{viola2}{cmyk}{0,1,0.2,0.6}
\definecolor{verde}{cmyk}{0.92,0,0.59,0.25}
\definecolor{verdec}{cmyk}{0.92,0,0.59,0.15}
\definecolor{verdes}{cmyk}{0.92,0,0.59,0.4}
\definecolor{verdino}{cmyk}{0.12,0,0.09,0.05}
\definecolor{giallo}{cmyk}{0,0,1,0}
\definecolor{gialloverde}{cmyk}{0.44,0,0.74,0}

\font\tenrsfs=rsfs10 at 12pt
\font\smallsfs=rsfs10 at 11pt
\font\sevenrsfs=rsfs7
\font\fiversfs=rsfs5
\newfam\rsfsfam
\textfont\rsfsfam=\tenrsfs
\scriptfont\rsfsfam=\sevenrsfs
\scriptscriptfont\rsfsfam=\fiversfs
\def\mathscr#1{{\fam\rsfsfam\relax#1}}
\def\Lag{\mathscr{L}}
\def\Lags{\hbox{\smallsfs L}}

\def\Amp{\mathscr{A}}

\begin{document}% IFUP-TH/2007-01
\color{black}
\vspace{0.5cm}
\begin{center}
{\Huge\bf\color{rossos}Thermal production of gravitinos}
\bigskip\color{black}\vspace{0.6cm}{
{\large\bf  Vyacheslav S. Rychkov}$^a$,
{\large\bf Alessandro Strumia}$^b$.
} \\[7mm]
{\it $^a$ Scuola Normale Superiore and INFN, Piazza dei Cavalieri 7, I-56126 Pisa, Italy}\\[3mm]
{\it $^b$ Dipartimento di Fisica dell'Universit{\`a} di Pisa and INFN, Italia}
\end{center}
\bigskip
\centerline{\large\bf\color{blus} Abstract}
\begin{quote}
We reconsider thermal production of gravitinos in the early universe,
adding to previously considered $2\to 2$ gauge scatterings:
a) production via $1\to2$ decays, allowed by thermal masses:
this is the main new effect;
b) the effect of the top Yukawa coupling;
c) a proper treatment of the reheating process.
Our final result behaves physically
(larger couplings give a larger rate) and is twice larger than previous results,
implying e.g.\ a twice stronger constraint on the reheating temperature.
Accessory results about
(supersymmetric) theories at finite temperature and gravitino couplings
might have some interest.

\color{black}
\end{quote}
{\small\tableofcontents}

\newpage

\section{Introduction}
We compute the abundance of gravitinos thermally produced in the
early universe at temperature $T$. In the usual scenario where
sparticles around the weak scale keep it naturally small, this
process implies an important constraint on the maximal reheating
temperature, possibly saturated if such gravitinos are all observed
Dark Matter (DM). If instead sparticles exist much above the weak
scale, gravitino production is one of their very few experimental
implications that survive.

%Experimental constraints on this scenario become strong enough to challenge its plausibility~\cite{GRS}.
%Experimental constraints have grown such that doubts...
%and string anthropic also allow sparticles
%around $M_Z$, or around $4\pi M_Z$, or at the string scale (which might be anywhere below the Planck scale)
%or anywhere, or split, or super-split.

%by demanding
%that the produced gravitinos do not destroy Big Bang Nucleosynthesis,
%see~\cite{Moroi} for a recent precise study.
%This constraint also applies if sparticles are too heavy to be seen at colliders.

\begin{figure}[t]
\begin{center}
$$\includegraphics[width=0.7\textwidth]{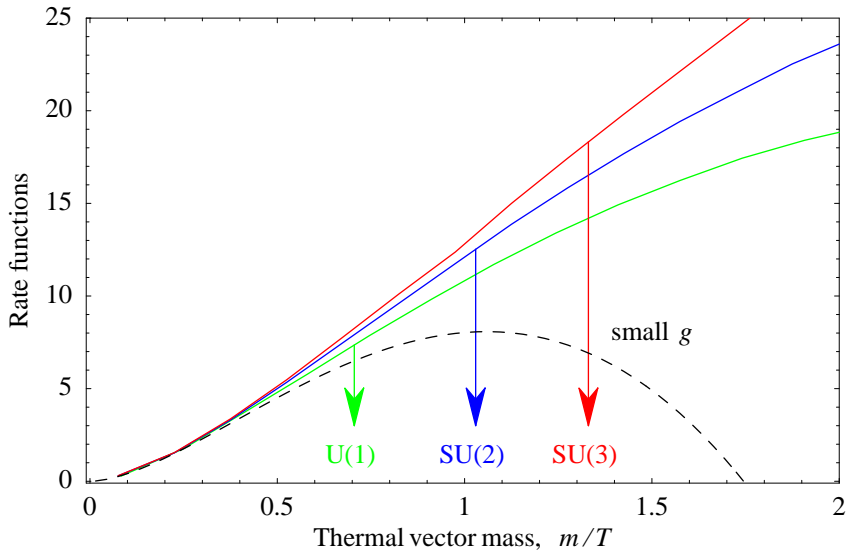}$$
\caption{\label{fig:res}\em Functions $f_3$, $f_2$ and $f_1$
that, as summarized in section~\ref{fNsummary}, 
describe our result for the gravitino production rate
from $\SU(3)_c$ (upper continuous curve, in red), $\SU(2)_L$ (middle continuous curve, in blue), ${\rm U}(1)_Y$ (lower continuous curve, in green) gauge interactions.
The arrows indicate the MSSM values of the thermal mass at $T\sim 10^{9}\GeV$.
The lower dashed curve shows the result from~\cite{Buch}, which agrees with our result in the limit of small
gauge coupling, and behaves unphysically for relevant ${\cal O}(1)$ values of the MSSM gauge couplings.
%The upper dashed curve indicates that the Hard Thermal Loop approximation is not accurate for this computation.
%\xxx{FISSARE $g\to 0$}
}
\end{center}
\end{figure}

%Since thermal production of gravitinos already has a vast literature, we start
%presenting the difference with respect to previous computations.
The gravitino production thermal rate was previously computed
in~\cite{gravitinoCosmo,Buch} at leading order in the gauge
couplings $g_3$ (and $g_2,g_Y$ in~\cite{postBuch}; we will add
effects due the top Yukawa coupling, which also has a sizeable
value). This roughly amounts to compute $2\to 2$ scatterings (like
gluon + gluon $\to$ gluon $\to$ gluino + gravitino), with thermal
effects ignored everywhere expect in the propagator of the virtual
intermediate gluon: a massless gluon exchanged in the $t$-channel
gives an infinite cross-section because it mediates a long-range
Coulomb-like force; the resulting logarithmic divergence is cut off
by the thermal mass of the gluon, $m\sim gT$, leaving a $\ln T/m$.
The explicit expression for the number of scatterings per space-time
volume, at leading order in the dominant QCD gauge coupling, was
found to be~\cite{Buch,postBuch}\footnote{Since we will adopt a
different technique, we cannot resolve the minor disagreement
between the results of~\cite{Buch} and~\cite{postBuch}.
%Anyhow we will find a significantly different result.
Notice also that, for later convenience, in eq.\eq{Buch} we explicitly show the power $\pi^5$
(following from the phase space for scattering processes, and
dictated by na\"{\i}ve dimensional analysis), which is explicitly
present in~\cite{gravitinoCosmo} and partially hidden in numerical coefficients in~\cite{Buch}.
}
\beq\label{eq:Buch}
 \gamma_{\rm scattering} =\frac{T^6}{2\pi^3 \bar M_{\rm Pl}^2}\left(1+\frac{M_3^2}{3 m_{3/2}^2}\right)f(g_3),\qquad
 f(g_3) = \frac{320.}{\pi^2}
g_3^2 \ln \frac{1.2}{g_3}\eeq
%\beq\label{eq:Buch}
% \gamma_{\rm scattering} =7.3 \frac{3\zeta(3)T^6}{16\pi^5 \bar M_{\rm Pl}^2}\left(1+\frac{M_3^2}{3 m_{3/2}^2}\right)
%g_3^2 \ln \frac{1.2}{g_3}\eeq
where $\bar M_{\rm Pl} = 2.4~10^{18}\GeV$ is the reduced Planck mass,
$M_3$ is gluino mass and $m_{3/2}$ is the gravitino mass.
This production rate unphysically decreases for $g_3\circa{>} 0.7$ becoming
negative for $g_3\circa{>}1.2$.
Fig.\fig{res} shows that the physical value, $g_3\approx 0.85$ at $T\sim 10^{10}\GeV$,
lies in the region where the leading-order rate function $f(g_3)$ (dashed line) is unreliable. 
Fig.\fig{res} also illustrates our final result (to be precisely described in section~\ref{fNsummary}): 
$f$ will be replaced by the continuous lines,
which agree with the leading order result at $g \sim m/T \ll 1$
and differ at $g\sim 1$.

% the function $f$

%As illustrated in fig.\fig{res} (to be precisely described in section~\ref{fNsummary})
% the physical value, $g_3\approx 0.85$ at $T\sim 10^{10}\GeV$, lies in the region where eq.\eq{Buch}
% (i.e.\ the dashed `small $g$' line in the plot) is unreliable.
% We will compute the function that replaces $g^2 \ln 1.2/g$ (valid for $g\ll1$)
% at physical values of the gauge coupling:
% the continuous lines in fig.\fig{res} basically show our result.

\medskip

%[

%In agreement with na\"{\i}ve dimensional analysis, the
%result for the number of scatterings per space-time volume at leading order in $g$
%has the form
%\beq \gamma_{\rm scattering} \sim\frac{T^6}{\pi^5 \bar M_{\rm Pl}^2}   g^2 \ln\frac{1}{g} \eeq
%where
%We now explain why this result is unsatisfactory.
%The practical reason can be easily seen looking at t

%

Let us now explain why the leading-order approximation in\eq{Buch} starts to be
inadequate already at $g\sim 0.7$. In thermal field theory higher
order corrections are usually suppressed by $g/\pi$: somewhat worse
than the usual expansion coefficient $(g/\pi)^2$ at $T=0$, but still
typically good enough at $g\sim 0.7$. Na\"{\i}ve power counting fails
(without signaling a breaking of the perturbative expansion) when
some new phenomenon only starts entering at higher orders, and this
is what happens in the case of gravitino production: a new simpler
process gives corrections of relative order $(g\pi)^2$.
%Gravitino
%production computations of \cite{Buch,postBuch} were modeled after
%the computation of axion production from QED plasma \cite{BY}.
%However, while the axion couples to two photons,
The gravitino couples to two particles with different thermal
masses: gluon/gluino, and quark/squark. 
Since thermal masses grow like $T$, this gives rise to a new
process with a rate growing like $T^6$: gravitino
production via decays, such as gluon $\to$ gluino + gravitino, whose
rate can be crudely estimated as \beq \gamma_{\rm decay} \sim
\frac{m}{T} \frac{T^3 \Gamma}{\pi^2} \sim \frac{m^4T^2}{\pi^3 \bar
M_{\rm Pl}^2} \sim \frac{g^4}{\pi^3}\frac{T^6}{\bar M_{\rm
Pl}^2}\eeq Indeed $\gamma_{\rm decay}$ is of course proportional to
the decay rate at rest $\Gamma\sim m^3/\pi \bar M^2_{\rm Pl} $;
which is slowed down by the Lorentz dilatation $m/T$ factor; the
$T^3$ takes care of dimensions, and less $\pi$ are present at the
denominator because a $1\to 2$ decay involves less particles than a
$2\to 2$ scattering.
%Inserting a typical thermal mass $M\sim gT$ one finds
%\beq \gamma_{\rm decay} \sim \frac{g^4}{\pi^3}\frac{T^6}{M_{\rm Pl}^2}\qquad\hbox{and so}\qquad
%\frac{\gamma_{\rm decay}}{\gamma_{\rm scattering}} \sim g^2 \pi^2\eeq
So, despite being higher order in $g$, the decay rate can be enhanced by a phase space factor $\pi^2$.
Subsequent higher order corrections should be suppressed by the usual $g/\pi$ factors.
Our goal is including such enhanced higher order terms, and this finite-temperature computation is
practically feasible  because a decay is a simple enough process.

\medskip

So far we explained the physical picture in a simple intuitive way.
A more precise  technical language is necessary to present how we
will proceed. To get the gravitino production rate we actually
compute the imaginary part of the gravitino propagator in the
thermal plasma. Thermal effects distort the dispersion relations
$E(k)$ of gluons, gluinos, quarks, squarks by i) adding a thermal
mass $E^2 = k^2 + m^2(k)$ to the modes already existing at  zero
temperature; ii) by introducing  new collective excitations (gluons
with longitudinal polarization, gluinos with `wrong' helicity, ...)
with their own dispersion relation; iii) beyond the two poles
mentioned above, the spectral densities of particles in a thermal
plasma also develop a `continuum' contribution, that can be thought
of as a parton-like distribution, with a continuum range of masses.
Physically it arises because particles can exchange energy with the
plasma.

%The `continuum' in iii) gives a `decay' contribution even in the
%axion case, $\gamma\to a \gamma$, and

In previous works~\cite{Buch,postBuch} the gluon thermal mass was taken into account
to regulate infra-red divergences encountered in scattering rates,
and the contribution of the gluon `continuum' was computed using
a standard technique introduced in axion computations \cite{BY},
that allows to extract the rate at leading order in $g$.
This was achieved by introducing an arbitrary splitting scale $k$ that obeys
$gT \ll k \ll T$.

\begin{figure}[t]
\begin{center}
$$\includegraphics[width=0.98\textwidth]{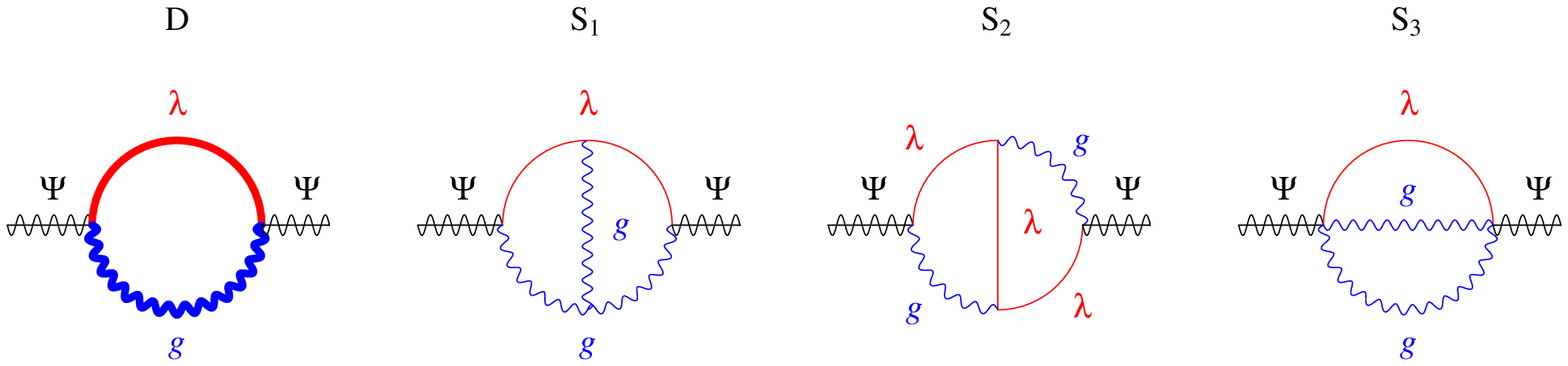}$$\vspace{-1cm}
\caption{\label{fig:Feyn1}\em Some Feynman diagrams that contribute to the imaginary part of the gravitino propagator. Thick lines denote resummed thermal propagators for the gluon $g$ and gluino $\lambda$. We do not plot diagrams involving quarks $q$ and squarks $\tilde{q}$, but they are of course included in our computation.}
$$\includegraphics[width=0.8\textwidth]{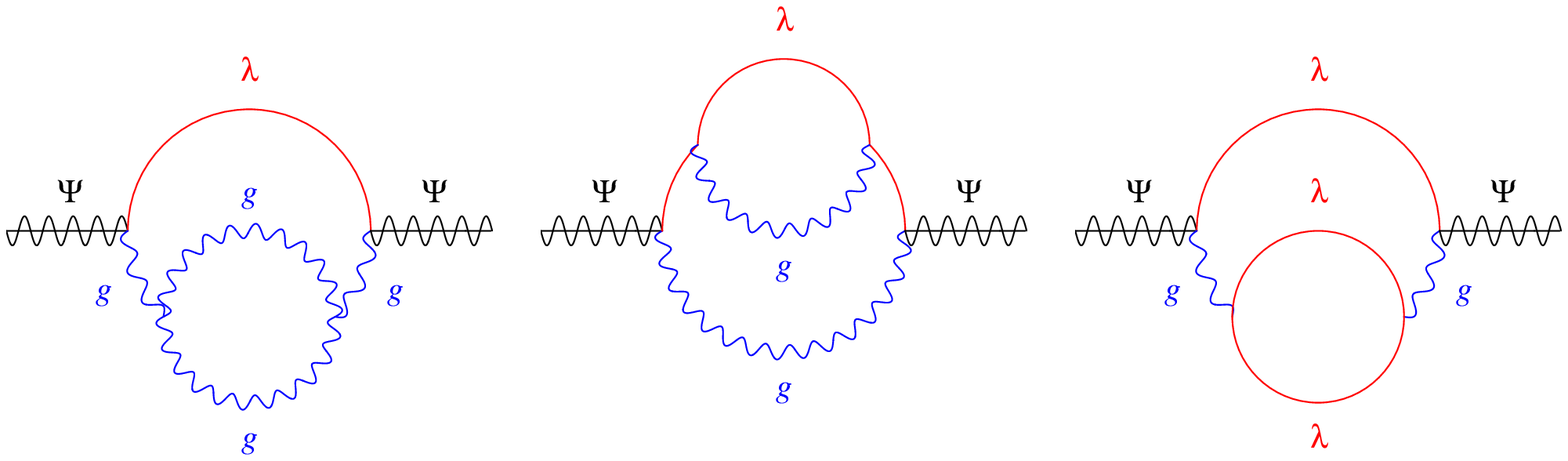}$$
\vspace{-1cm}
\caption{\label{fig:Feyn3}\em Two-loop Feynman diagrams that appear in the expansion of  diagram ${\rm D}$, that resums
all higher loop diagrams with iterated one-loop corrections to
gluon and gluino propagators.}
\end{center}
\end{figure}

We will not use this technique:
because its validity is doubtful for  $g_3 \approx 0.85$,
and because we actually
want to include the enhanced higher order terms,
taking into account that a gravitino (unlike an axion) couples to two particles
with different thermal masses.
We will instead compute the decay diagram (D in fig.\fig{Feyn1}) using resummed finite-temperature propagators for gluons, gluinos, quarks, squarks.
The perturbative expansion of this diagram D contains the two-loop diagrams in  fig.\fig{Feyn3}:
their imaginary parts correspond to well-defined combinations of scattering processes, as dictated by cutting rules.
This fixes how scatterings must be subtracted in order to avoid overcountings of effects already
 described by thermal masses via diagram D.
 In section~\ref{subtractions} we compute the subtracted scattering rates,
 in section~\ref{decay} we compute the gravitino production rate via `decay',
 and in section~\ref{top} we add the rate due to the top quark Yukawa coupling.

In section~\ref{reh} we sum these effects and compute the gravitino
abundance writing a set of Boltzmann equations that describe the
reheating process, previously approximated assuming a
maximal temperature equal to the reheating temperature $T_{\rm RH}$.
%(previously approximated with the universe
%suddenly appearing at a maximal temperature $T_{\rm RH}$).
Our
results are summarized in the conclusions,
section~\ref{conclusions}.

In the passing we address some issues related to finite-temperature
and to supersymmetry. In section~\ref{TQFT} we list explicit values
for thermal masses for all particles and sparticles, noticing that
they obey some supersymmetric relation. Appendix~\ref{gravitino}
gives a (non uselessly) fully precise summary of gravitino
interactions, and in appendices~\ref{thermal}, \ref{thermalF} we
collect full expressions for the thermal corrections to vector and
fermion propagators.

\begin{figure}[t]
\begin{center}
$$\includegraphics[width=0.9\textwidth]{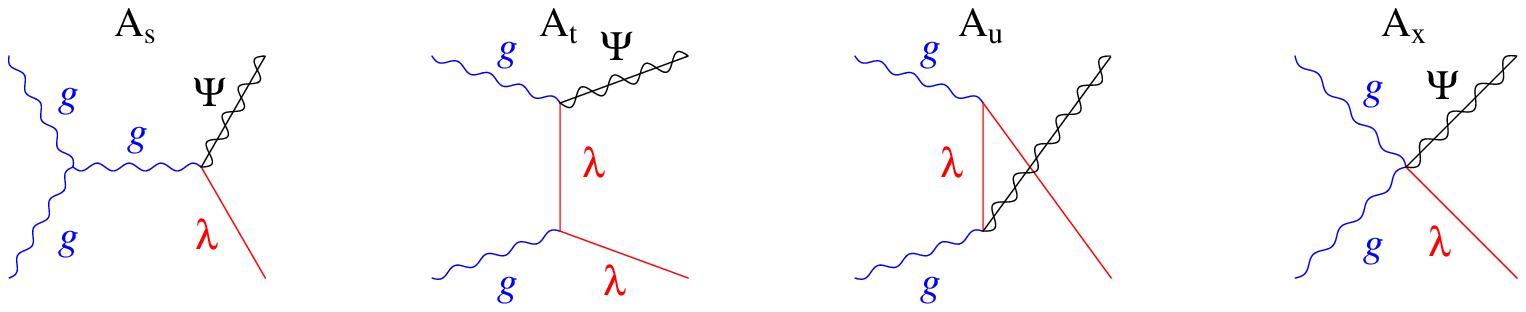}$$
\caption{\label{fig:FeynA}\em Feynman diagrams that contribute to $gg\to \lambda\gr $ scatterings.}
$$\includegraphics[width=0.9\textwidth]{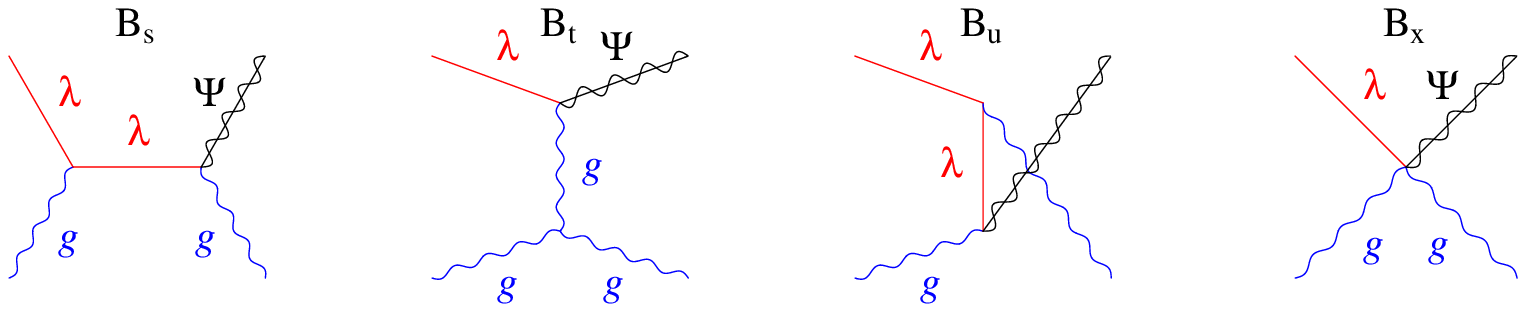}$$
\caption{\label{fig:FeynB}\em Feynman diagrams that contribute to $g\lambda\to g\gr$ scatterings.}
%the gravitino production rate.
%Analogous diagrams ${\rm A}_{\rm s}$, ${\rm B}_{\rm s}$, ${\rm C}_{\rm s}$, ${\rm A}_{\rm u}$, ${\rm B}_{\rm u}$, ${\rm C}_{\rm u}$ (not plotted) have the intermediate particle exchanged in the $s$ and $u$ channels. }
\end{center}
\end{figure}

\setcounter{equation}{0}

\section{Subtracted scattering rate}\label{subtractions}
Gravitinos $\Psi_{\mu}$ with momentum
$P=(E,\mb{p})$ are
produced via their coupling $\bar{\Psi}_{\mu}S^{\mu}/{2\bar{M}_{\mathrm{Pl}}}%
$, where $S^{\mu}$ is the supercurrent of the visible sector of a
supersymmetric theory, here assumed to be the MSSM. The visible
sector is thermalized, while the gravitino is not, since its
coupling to the MSSM plasma is weak. According to the general
formalism of thermal field theory \cite{LeBellac}, the production
rate of such a weakly interacting fermion is related to the
imaginary part of its propagator $\Pi$ as
\begin{equation}
\gamma= \frac{dN}{dV\,dt}=-2\int d\vec{P}\,f_{F}(E)\Im\Pi=\int d\vec{P}%
~\Pi^{<}(P),\qquad d\vec{P}\equiv\frac{d^{3}p}{2E(2\pi)^{3}} .
\label{eq:ImProp}%
\end{equation}
Here $\Pi^{<}$ is the non time-ordered gravitino propagator summed
over its polarizations i.e.\ traced with the gravitino polarization
tensor $\Pi_{\mu \nu}$ (appendix~\ref{gravitino} gives explicit
expressions):
\begin{equation}
\Pi^{<}(P)=\frac{1}{4\bar{M}_{\mathrm{Pl}}^{2}}\Tr\big[\Pi_{\mu\nu}(P)\langle
S^{\nu}(P){\bar{S}}^{\mu}(-P)\rangle_{T}\big] \label{eq:Pi<}%
\end{equation}
where $\langle\cdots\rangle_{T}$ denotes thermal average. We employ
$\Pi^{<}$ because it gives slightly cleaner formul\ae {} than
$\Im\Pi$. Eq. (\ref{eq:ImProp}) is valid at leading order in the
gravitino coupling $\bar {M}_{\mathrm{Pl}}^{-1}$, and to all orders
in the MSSM couplings, $g_{Y,2,3}$ and $\lambda_{t}$. Extracting
predictions from (\ref{eq:ImProp}) is limited only by our ability to
evaluate (\ref{eq:Pi<}).

Thermal field theory cutting rules allow to see that, at leading
order in the MSSM couplings, eq.\eq{ImProp} is equivalent to summing
rates for the various tree-level processes that lead to gravitino
production. At tree level this formalism is more cumbersome than a
direct computation of production rates. However, in this paper we
want to take into account finite temperature corrections to the MSSM
particle propagators arising at one loop level: eq.\eq{ImProp}
becomes more convenient because it cleanly dictates how one must
resolve ambiguities encountered in scattering computations that
arise because Lorentz invariance is broken by the thermal plasma.
%(more later).

\smallskip

Fig.\fig{Feyn1} shows some main Feynman diagrams that contribute to
$\Im\Pi$. What we actually compute in this paper is the first
one-loop diagram `D', using the Feynman gauge resummed
finite-temperature propagators for the gluon and gluino in the loop.
Therefore it describes a sum of an infinite number of multi-loop
diagrams: the lowest-order ones are shown in fig.\fig{Feyn3}.
Resummation is needed because thermal effect drastically change the
gluon and gluino propagators, in particular opening a phase space
for decays, such as $g\rightarrow\lambda\Psi$ and/or
$\lambda\rightarrow g\Psi$. Clearly diagram D contains this decay
process. However, by cutting fig.\fig{Feyn3}, one sees that diagram
D also describes some of the $2\rightarrow2$ scattering processes
computed in previous analyses~\cite{Buch}. Therefore, before
starting the computation, we clarify this issue showing how the
total gravitino production rate is obtained.

\medskip

The total scattering rate is the sum of various $2\to2$ processes A,
B, C,\ldots, listed in table~\ref{tab:diffcs}. Each process is the
modulus squared of the sum of a few amplitudes, corresponding to the
single Feynman diagrams, that we label as $s,t,u,x$:
\[
\gamma_{\mathrm{scattering}} = |A_{s} + A_{t}+A_{u}+A_{x}|^{2} +
|B_{s} + B_{t} + B_{u}+B_{x}|^{2} + \cdots.
\]
This notation indicates that often 4 diagrams contribute to a given
process: 3 diagrams are generated by $s$, $t$ and $u$-channel
exchange of some particle among two vertices ($g_{3}$ and
$1/M_{\mathrm{Pl}}$), and a fourth diagram arises from a quartic
supergravity vertex with coupling $g/M_{\mathrm{Pl}}$.
Fig.\fig{FeynA} and\fig{FeynB} show concrete examples of the 4
diagrams that contribute to $gg\to\Psi\lambda$ and to
$g\lambda\to\Psi g$ scatterings, respectively. The latter rate is
logarithmically infra-red (IR) divergent, because diagram B$_{t}$ is
mediated by $t$-channel gluon exchange, that describes a
Coulomb-like scattering.

The main result can be obtained by careful visual inspection of
cutting rules: diagram D describes the sum $|A_{s}|^{2} +
|A_{t}|^{2} + |A_{u}|^{2} + |B_{s}|^{2} + |B_{t}|^{2}
+|B_{u}|^{2}+\cdots$ of the modulus squared of all $2\to2$ diagrams
that contain the gauge coupling $g_{3}$. $2\to2$ scattering rates
generated by supergravity quartic vertices are instead described by
diagrams like S$_{3}$ in fig.\fig{Feyn1}. Some cuts of the two loop
diagrams like S$_{1}$ and S$_{2}$ describe the interference terms
among the various Feynman diagrams.
%${\rm S}_1 = {\rm A}_{\rm s}{\rm A}_{\rm t}^*+{\rm B}_{\rm s}^*{\rm B}_{\rm t}+\cdots$
%{\bf CHECK}
(Notice that the imaginary part of a single two-loop diagram can
describe contributions to different scattering processes). Other
cuts of these diagrams describe one loop corrections to the
gravitino vertices, that do not give any leading order contribution
if thermal masses are neglected. Thermal masses open a phase space
for $1\to2$ processes (we can neglect the decay rate generated
by zero-temperature masses $m$, since we are interested in $T\gg m$), 
and we will later
argue that we can still neglect thermal corrections to the gravitino
vertex.

\begin{table}[h]
\begin{center}%
\begin{tabular}
[c]{c|r@{~$\to$~}l|ccc}
& \multicolumn{2}{c|}{process} & $|\Amp|^{2}_{\mathrm{full}}$ & $|\Amp|^{2}%
_{\mathrm{subtracted}}$ & \\\hline\hline F & $\gl \gl $ & $\gl \gr $
& $-8C{(s^{2}+t^{2}+u^{2})^{2}}/{stu}$ & 0 &
\\\hline
A & $g g $ & $\gl \gr $ & $\phantom{+}4C(s+2t+2t^{2}/s)$ & $-2sC$ &
\\\hline B & $g \gl $ & $g \gr $ & $-4C(t+2s+2s^{2}/t)$ &
$\phantom{+}2tC$ & \\\hline H & $\sq \gl $ & $\sq \gr $ &
$-2C^{\prime}(t+2s+2{s^{2}}/{t})$ & $-tC^{\prime}$ & \\\hline
J & $\sq \bar{\sq} $ & $\gl \gr $ & $\phantom{+}2C^{\prime}(s+2t+2{t^{2}}%
/{s})$ & $\phantom{+}sC^{\prime}$ & \\\hline C & $\sq g $ & $q \gr $
& $\phantom{+}2sC^{\prime}$ & 0 & \\\hline D & $g q $ & $\sq \gr $ &
$-2tC^{\prime}$ & 0 & \\\hline E & $\bar{\sq} q $ & $g \gr $ &
$-2tC^{\prime}$ & 0 & \\\hline G & $q \gl $ & $q \gr $ &
$-4C^{\prime}(s+{s^{2}}/{t})$ & 0 & \\\hline I & $q \bar{q} $ & $\gl
\gr $ & $-4C^{\prime}(t+{t^{2}}/{s})$ & 0 & \\\hline
\end{tabular}
\medskip
\end{center}
\caption{\textit{Squared matrix elements for gravitino ($\gr$)
production in units of $g^{2}_{N}/\bar{M}^{2}_{\mathrm{Pl}} (1+
M^{2}_{N}/3m_{3/2}^{2})$, summed over all polarizations and gauge
indices. The result hold for all three factors of the SM gauge group
with $N=\{1,2,3\}$, although the notations are appropriate for the
$\SU(3)$ case: $g,\lambda, q,\tilde{q}$ denote gluons, gluinos,
quarks, squarks. The gauge factors $C_{N}$ and $C^{\prime}_{N}$ are
defined in the text. Rates A and B are the sum of vector and ghost
contributions. }}%
\label{tab:diffcs}%
\end{table}

In conclusion, the total gravitino production rate due to gauge
couplings will be computed as
\begin{equation}
\gamma=\gamma_{D}+\gamma_{S}^{\mathrm{sub}},
\end{equation}
the sum of diagram D (that describes decay plus modulus squared of
many single $2\rightarrow2$ diagrams) plus the set of remaining
$2\rightarrow2$ rates, obtained by subtracting from the total
scattering rate $\gamma_S$ the effects already included in
$\gamma_D$.
Explicit results for $\gamma_D$ and for $\gamma_S^{\rm sub}$
will be given in eq.\eq{res} and eq.\eq{Ssub} respectively,
and the conclusions will describe how to use them.

%\subsection{Equivalence theorem and gravitino gauge invariance}

\bigskip

Before proceeding to actual computations, we have to clarify the
issues of gravitino coupling and gravitino gauge invariance. We are
interested in $T\gg m$, where $m$ denotes sparticle or gravitino
masses: gravitino $\leftrightarrow$ Goldstino equivalence
(appendix~\ref{gravitino}) means that at leading order in $m/T$ the
massive gravitino field $\Psi_{\mu}$ can be replaced with two
massless field: a massless gravitino $\psi$ coupled to the
supercurrent $S_{\mu}$ (given in eq.\eq{4-comp}, it can be evaluated in
the supersymmetric limit ignoring soft terms) plus a massless
Goldstino $\chi$, coupled to the divergence of the supercurrent
(given in eq.\eq{dS}, only the soft terms factored out are
relevant):
\begin{equation}
\Lag_{\mathrm{int}}=\frac{\bar{\psi}_{\mu}S^{\mu}}{2\bar{M}_{\text{Pl}}}%
+\frac{\bar{\chi}\,(\partial_{\mu}S^{\mu})}{\sqrt{6}\bar{M}_{\mathrm{Pl}%
}m_{3/2}}\ . \label{eq:ET}%
\end{equation}
The gravitino production rate is given by
$\gamma(\Psi_{\mu})\simeq\gamma (\psi_{\mu})+\gamma(\chi)$.
%The supersymmetry will be broken explicitly by soft terms and $\partial S\sim
%m_{\text{soft}}$ is given in the Appendix. Computation of the Goldstino
%production rate is thus unambiguous. The massless gravitino production rate as
%given by (\ref{eq:equiv}) is ambiguous due to this current nonconservation,
%however the ambiguity is negligible and can be ignored as we will now explain.
%The way to study this ambiguity is to see how the production rate changes when
%the projector of massless gravitino is changed by gravitino gauge
%transformation. For instance one can compare the transverse, the van N and the
%"physical" projectors. We see that the projector changes by terms proportional
%to $P_{\mu}O(1)$ which will give production rate suppressed by a factor
%$m_{\text{soft}}/T$ with respect to the inambiguous terms. The conclusion is
%that the soft terms are relevant only for the Goldstino production, but can be
%ignored when computing the massless gravitino production rate.
%The above ambiguity $O(m_{\text{soft}}/T)$ is inavoidable but totally
%negligible.
While the total rate is gauge independent (vectors have $\SU(3)_{c}%
\otimes\SU(2)_{L}\otimes\mathrm{U}(1)_{Y}$ gauge invariance; the
computation of $\gamma(\psi_{\mu})$ also involves gravitino gauge
invariance), its splitting in resummed and not-resummed
contributions is not. We are resumming a well-defined class of
effects, but we cannot systematically include all the effects up to
a given order in $g$: therefore our result has a residual
gauge-dependence, of relative order $g^2/\pi^2$, due to partial
inclusion of higher-order terms.
%We will later discuss gauge-dependent
To make the computation feasible, we choose for vectors the Feynman
gauge, and for the massless gravitino $\psi_{\mu}$ the gauge where
its propagator and polarization tensor does not involve terms
containing $P_{\mu}$ or $P_{\nu}$, eq.\eq{PiSenzaP}:
\begin{equation}
\Pi_{\mu\nu}^{3/2}=-\frac{1}{2}\gamma_{\mu}P\hspace{-1.5ex}/\,\gamma_{\nu
}-P\hspace{-1.5ex}/\,\eta_{\mu\nu}. \label{eq:ripetuta}%
\end{equation}
One first motivation for this choice is that, in the supersymmetric
limit, the full supercurrent satisfies $P_{\mu}S^{\mu}=0$, while
sub-sets of $S_{\mu}$ are not separately conserved: with
choice\eq{ripetuta} we never have to deal with such terms. Of
course, the same gauge is used for computing both the resummed
diagram D and the subtracted scattering rates.

Table~\ref{tab:diffcs} gives explicit values for the subtracted
massless gravitino  and Goldstino
scattering rates due to gauge interactions. It is important to
notice that, unlike the total rate, \emph{the subtracted rates are
infra-red convergent}: no $1/t$ factors appear because all divergent
Coloumb-like scatterings, like $|B_{t}|^{2}$, are included in
diagram D, that we compute using thermal masses that provide the
physical cut-off. Unlike in the conventional technique~\cite{BY}
employed in~\cite{Buch,postBuch}, our technique does not need to
introduce an arbitrary splitting scale $k_{\ast}$ that satisfies the
problematic conditions $gT\ll k_{\ast}\ll T$ in order to control
infra-red divergences. Some contributions to subtracted scattering
rates turn out to be negative, but the total rate will be positive
and dominated by diagram D. In Feynman gauge, rates for the
processes A and B (the ones that involve two vectors) actually are
the sum of scatterings involving two vectors (four diagrams,
computed with the Feynman polarization tensor
$\sum\epsilon_{\mu}\epsilon_{\nu}^{\ast}=-\eta_{\mu\nu}$) plus
scatterings containing two ghosts (one diagram, negative
$|\Amp|^{2}$).

A curious fact happens. Despite the fact that the massless gravitino
$\psi$ and the Goldstino $\chi$ have different couplings (in
particular the Goldstino has no coupling to quark/squark, and
consequently a reduced set of Feynman diagrams), the differential
production cross sections for these two particles are the same,
process by process, up to the universal factor
$M_{N}^{2}/3m_{3/2}^{2}$, where $M_{1,2,3}$ are the gaugino masses.
We don't know if there is a simple generic reason behind this
equality. The second reason for choosing the gravitino projector of
eq.\eq{ripetuta} is that it respects this equality also for
subtracted scattering rates.

Subtracted rates for processes C, D, E, G, I vanish, and looking at
Goldstinos one can easily understand why: a single Goldstino diagram
contributes, such that no interference terms exist. This is not the
case for scatterings H and J, where a second Goldstino diagram
contributes, generated by the quartic Goldstino coupling in
eq.\eq{dS}. (This extra coupling is not present for the ghost
scatterings in A and B analogous to H and J, as we employ a
non-supersymmetric gauge without ghostinos). In case of scattering F
the subtracted rate vanishes because proportional to $s+t+u=0$. A
$1/2!$ factor must be included for the A and F processes that have
equal initial state particles, and a factor 2 for C, D, G, H that
can occur with particles and with anti-particles. The total result
for the subtracted gravitino production rate is
\begin{equation}
\gamma_{S}^{\mathrm{sub}}=1.29\frac{T^{6}}{8\pi^{5}\bar{M}_{\mathrm{Pl}}^{2}%
}\sum_{N=1}^{3}g_{N}^{2}(1+\frac{M_{N}^{2}}{3m_{3/2}^{2}})(C_{N}^{\prime
}-C_{N}) \label{eq:Ssub}%
\end{equation}
where the numerical factor accounts for the difference with respect to the
scattering
rate computed in Boltzmann approximation, where $\gamma\approx\sigma T^{6}%
/\pi^{4}$ where $\sigma=\sum\int_{-s}^{0}dt~|\Amp|^{2}/16\pi s^{2}$
is a
constant. The sum runs over the three components $\mathrm{U}(1)_{Y}%
\otimes\SU(2)_{L}\otimes\SU(3)_{c}$ of the MSSM gauge group with
$N=\{1,2,3\}$, and $C_{N}=|f^{abc}|^{2}=N(N^{2}-1)=\{0,6,24\}$ and
$C_{N}^{\prime}=\sum_{\Phi}|T_{ij}^{a}|^{2}=\{11,21,48\}$ where
$\sum_{\Phi}$ runs over all chiral multiplets. We use the standard
normalization for hypercharge, where left-handed leptons have
$Y=-1/2$, that differs from the SU(5) normalization by a factor
$\sqrt{3/5}$.
%pREVIOUS VERSIONS WITH DIFFERENT NOTATIONS
%era  gamma = (N^2-1) [  - N  +  \frac{N_f}{8} ]
%C'_N=  Nf (N_c^2-1)/2$.
%One has $N=\{3,2,0  OR 1 \}$ and $N_f=\{6,7,11\}$ for the three $\SU(3)_c \otimes\SU(2)_L\otimes U(1)_Y$
%factors of the SM gauge group $i=\{3,2,1\}$.
All parameters are renormalized at an energy scale $\mu\sim T$.

The next step is computing diagram D: we first need to introduce finite temperature effects.

\setcounter{equation}{0}
\section{Finite temperature effects}\label{TQFT}
We here summarize some well known results from quantum field theory at finite temperature
that are relevant for our computations: the spectral densities of scalars, fermions and vectors
that  play a r\^ole analogous to parton densities in hadron scattering processes.
This section also contains a few original points:
practical formul\ae{} for thermal masses that apply to generic supersymmetric  models,
the observation that thermal effects respect supersymmetry  at $E\gg T$;
we explain what qualitatively changes and why we must go beyond the
Hard Thermal Loop approximation;
we discuss a possibly non-standard point of view about the problem of
negative spectral densities.

\begin{figure}[t]
\begin{center}
$$\includegraphics[width=\textwidth]{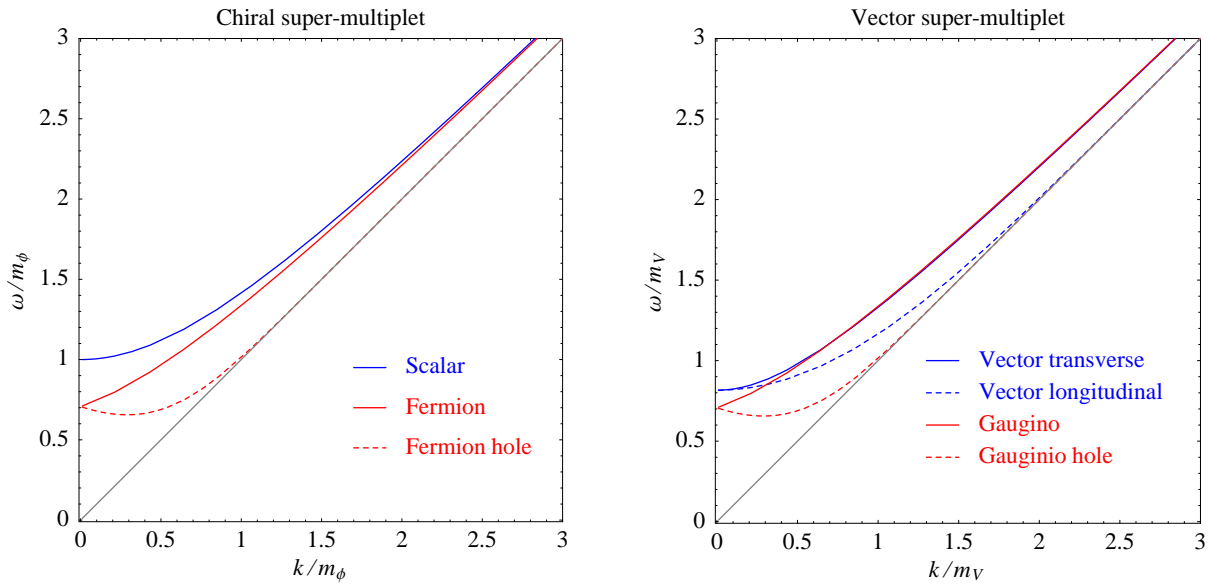}$$
\caption{\label{fig:SUSYMT}\em Dispersion relations at finite temperature in HTL approximation
for the components within a chiral (left) and vector (right) massless super-multiplet. Thermal effects are supersymmetric at $k\gg m\sim gT$.}
\end{center}
\end{figure}

\subsection{The Hard Thermal Loop approximation}
Thermal corrections simplify when one restricts the attention to diagrams with `{\em soft}' external momenta, $k\ll T$~\cite{Pisarski92,LeBellac}.
This approximation is useful if couplings are small, $g\ll 1$,
as it describes collective phenomena that develop at energies of ${\cal O}(gT)$
via simple effective thermal Lagrangians.
In the rest frame of the plasma, the non-local HTL Lagrangian for scalars $\phi$,
fermions $\psi$ and vectors is~\cite{Pisarski92,LeBellac}
\beq\label{eq:LHTL} \Lag_{\rm HTL} = m_S^2|A|^2 +
m_F^2 \int_\Omega   \bar\psi \frac{i\hat{K}\hspace{-1.5ex}/}{\hat{K}\cdot D} \psi -
m_V^2  {\rm Tr}\int_\Omega F_{\mu\alpha}\frac{\hat{K}_\alpha \hat{K}_\beta}{(\hat{K}\cdot D)^2}
F_{\beta\mu}+\cdots \eeq
where $\cdots$ denotes Yukawa or scalar couplings that do not receive HTL corrections;
gauge couplings receive thermal corrections such that $\Lag_{\rm HTL}$ is
gauge invariant (indeed $D$ denotes the usual gauge-covariant derivative);
$\hat K = (1,\hat{\mb{k}})$ is the `loop' momentum ($\hat{K}^2=0$);
$\int_\Omega = \int d\Omega/4\pi$ denotes angular average.
It is performed analytically in the more explicit results in
appendices~\ref{thermal} and~\ref{thermalF}

The key parameters are `thermal masses' of order $m\sim gT$. By
explicit computation we find the following values for thermal masses
in an unbroken supersymmetric theory with massless chiral
$\Phi=(\phi,\xi )$ and vector $V=(V_\mu,\lambda)$
superfields:\footnote{$\phi$ is a complex scalar, $\xi$ and
$\lambda$ are Weyl fermions. Explicit formul\ae{} for thermal masses
of bosonic sparticles had been given in~\cite{CE}; we agree with
their results.}
\beq\label{eq:MTSUSY} m_\phi^2= 2{m_\xi^2}=
\bigg[\frac{C_R}{2} g^2 + \frac{1}{4}\lambda^2\bigg]T^2, \qquad
m_V^2 = 2 m_\lambda^2 = \bigg[g^2\frac{C_V + T^2_R}{4}\bigg] T^2
\eeq where $g$ is the gauge coupling and $\lambda$ the coupling in
the superpotential $W = \lambda \Phi \Phi' \Phi''$. Summation over
gauge, flavor and any indices is understood. The group factors $C_R$
and $T^2_R$ are defined as ${\rm Tr}T_R^a T_R^b = T_R^2 \delta^{ab}$
(index of the representation) and as $(T_R^a T_R^a)_{ij}=C_R
\delta_{ij}$ (quadratic Casimir) where the generators are in the
representation $R$. By summing both over $ij$ and over $ab$ one
finds that they are related by $T^2_R \dim G = C_R \dim R$. Explicit
values are $T_R^2=1/2$ and $C_R = (N^2 - 1)/2N$ for the fundamental
of SU($N$) ($\dim R = N$, $\dim G = N^2-1$), $C_V=N$ for the adjoint
of SU($N$), and $C_R= q^2$ for a representation of U(1) with charge
$q$. In the MSSM with $3$ generations and one pair of Higgses one
has the following vector thermal masses \beq m_{V_3}^2
=\label{eq:mVn}
 \frac{9}{4} g_3^2 T^2,\qquad
 m_{V_2}^2 =
  \frac{9}{4} g_2^2 T^2, \qquad
m_{V_1}^2 = \frac{11}{4} g_Y^2 T^2\eeq
and the following scalar masses
\beq
m_{\tilde{E}}^2 = \frac{g_Y^2}{2}  T^2,\qquad
m_{\tilde{L}}^2 = m_{H_{\rm d}}^2 = \bigg[\frac{3}{8} g^2_2 + \frac{g^2_Y}{8} \bigg] T^2,\qquad
m_{H_{\rm u}}^2 = \bigg[\frac{3}{8} g^2_2 + \frac{g^2_Y}{8} +\frac{3}{4}\lambda_t^2 \bigg]T^2\eeq
$$
m_{\tilde{Q}}^2 =\bigg[\frac{2}{3} g_3^2 +  \frac{3}{8} g^2_2 + \frac{g^2_Y }{72} +
\frac{ \lambda_t^2}{4}\bigg] T^2 ,\qquad
m_{\tilde{U}}^2 = \bigg[\frac{2}{3} g_3^2 +  \frac{2}{9} g^2_Y + \frac{\lambda_t^2 }{2}\bigg] T^2,\qquad
m_{\tilde{D}}^2 = \bigg[\frac{2}{3} g_3^2 +  \frac{g^2_Y}{18}   \bigg] T^2
$$
where the $\lambda_t$ terms are present only for third generation squarks, and we
neglected analogous $\lambda_b$ and $\lambda_\tau$ terms, possibly relevant if $\tan\beta \sim m_t/m_b$.
Squared thermal masses for gauginos, higgsinos, quarks and leptons are a factor 2 smaller, as
summarized in\eq{MTSUSY}.

\medskip

We followed the standard convention for thermal masses.
Let us recall how they parameterize thermal dispersion relations $\omega(k)$ where
$\omega$ and $k$ are the energy and momentum with respect to the plasma rest frame.
Scalar thermal masses $m^2$ correspond to the relativistic dispersion relation
$\omega^2 = k^2 + m^2$, see eq.\eq{LHTL}.
For fermions the thermal mass $m$ tells the energy at rest of particle and hole (or `plasmon')
excitations, $\omega(k=0)=m$, while at large momentum the hole disappears\footnote{More precisely,
its residue at the pole is exponentially suppressed by $k^2/m^2$.
The fact that residues $Z(k)$ are not constant is one reason why
computing the imaginary part of the gravitino propagator in terms of particle and sparticle
spectral densities is a better formalism  than directly computing the gravitino production rate:
it precisely dictates how all these non-relativistic factors must be taken into account.}
and particles have $\omega^2(k\gg T) \simeq k^2 + 2 m^2$.
For vectors the thermal mass $m$ tells the dispersion relation of transverse polarizations
at large momentum,
$\omega^2(k\gg T) \simeq k^2 + m^2$, while at rest both transverse and
longitudinal polarizations have energy $\omega^2(k=0)=2 m^2/3$.

Therefore, despite the misleading conventional factors 2,
eq.\eq{MTSUSY} means that
{\em within each multiplet, vector or chiral, thermal effects  at $k\gg T$
modify in the same way the dispersion relation of its bosonic and of its fermionic components}.
This likely is a consequence of the eikonal theorem,
that tells that gauge interactions with soft vectors
do not depend on the particle spin but only on its gauge current
(thermal masses physically describe the kinetic energy that
a particle acquires due to scatterings with the thermal plasma).
Fig.\fig{SUSYMT} shows the dispersion relations $\omega(k)$
of the particles within chiral and vector multiplets.
Particles and sparticles have similar dispersion relations,
reducing the phase space for gravitino production via decays.

\begin{figure}[t]
\begin{center}\vspace{-1cm}
$$\includegraphics[width=0.8\textwidth,height=0.5\textheight]{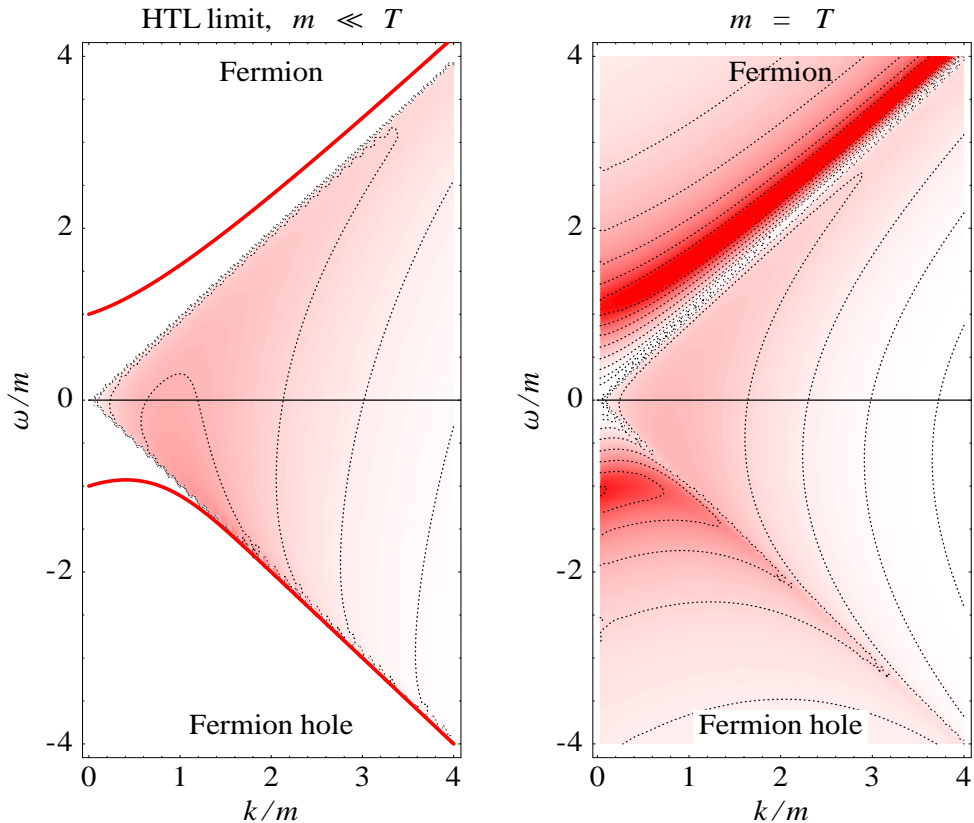}$$
%$$\includegraphics[width=0.8\textwidth,height=0.5\textheight]{Vectorm}$$
\vspace{-16mm}
\caption{\label{fig:Fermionm}\em Spectral density of a massless fermion in a thermal plasma,
plotted in HTL approximation ($g\ll 1$, i.e.\ thermal mass $m\sim gT \ll T$) and beyond.
Notice the main differences:
the particle ($\omega>0$) and `holes' ($\omega<0$) poles develop a finite width, and,
more importantly,
the continuum below the light-cone gets
Boltzmann suppressed at $k\circa{>} T$.
Contours are equispaced in $\log$ scale.
}
\end{center}
\end{figure}

\subsection{Full one loop thermal effects}
The HTL approximation holds at momenta and energies  $k,\omega \ll T$,
correctly describing thermal effects that arise at $k,\omega \sim gT$ if $g\ll 1$.
However, the physically relevant values of gauge couplings
(especially the strong coupling) are not small enough to justify the use
of the HTL approximation.
We therefore use the full one-loop thermal and quantum corrections to propagators of
scalars, fermions and vectors.
Explicit expressions are collected in appendices~\ref{thermal} and~\ref{thermalF}
(see also~\cite{WeldonFermion,WeldonVector,thoma,LeBellac})
and fig.\fig{Fermionm} illustrates (in the case of a massless fermion)
the qualitatively new effects that arise beyond the HTL limit.

The most visible effect (although not the most important one) happens at $|\omega|>k$
i.e.\ `above the light cone'.
In the $g\to 0$ limit particles (and quasi-particles such as fermion `holes')
have an infinitesimal width: their dispersion relations are plotted as
thin lines in fig.\fig{Fermionm}a.
For finite $g$ they get a finite width $\Gamma$ (both from $T=0$ quantum effects and from
thermal effects), such that their spectral density
gets smeared acquiring the usual bell-like shape.
This is why a continuum appears also above the light cone in fig.\fig{Fermionm}b.
A well known problem encountered in thermal computations
is that sometimes thermal effects give $\Gamma <0$.
We therefore included the $T=0$ contribution, finding that the total $\Gamma$ is positive
for scalars and fermions.
%In agreement with earlier studies, we find that fermions have $\Gamma>0$.
%In the simpler case of scalars we verified that the $T=0$ contribution
%to $\Gamma$ has exactly the minimal value needed to make the full $\Gamma$ positive
%for all values of the momentum. Of course, the physical $\Gamma$ is the full one.
This cure does not work for vectors, because the $T=0$ contribution
to their $\Gamma$ can itself   be negative depending on the gauge choice, see eq.\eq{pi0}.
%(and on if the non-abelian gauge contribution dominates over the matter contribution).
We therefore think that the negativity of $\Gamma$ is not related to higher-order subtleties in the
thermal expansion, but just to gauge invariance.
It should not affect computations of physical gauge-invariant quantities, provided
that one can do an exact computation up to some order in the perturbative expansion.
The only trouble is that in practice it is difficult to achieve this in finite temperature computations.
In view of this situation, since the would-be poles are anyhow reasonably narrow for
the physical values of the coupling that enter our computation,
we use for them the HTL approximation\footnote{
This might be not an entirely satisfactory approximation
for the pole-pole contribution to gravitino production, as
particle and sparticles happen to have similar dispersion relations at $k\sim T$,
and what matters for the phase space is their mass difference.
Due to this reason, we will find that the pole-pole contribution is small,
and it seems unlikely that adding a finite width can change this conclusion.

Furthermore, one might compensate this approximation by not subtracting
modulus squared of $s$-channel diagrams when computing subtracted rates.
Since these details have negligible numerical significance, we prefect to avoid them.}.
Notice that the HTL approximation correctly describes the position of the poles
(i.e.\ the dispersion relations) even at large $k\circa{>}T$:
poles lie close to the light-cone, $|\omega|\approx k$, even if $g\sim 1$~\cite{thoma}.
%The longitudinal and hole collective excitations have a residue that exponentially vanishes
%at $p \gg m$, therefore there is no need of accurately approximating their dispersion relations
%outside the soft regime.

\bigskip

The new effect important for our purposes arises at $|\omega|<k$,
i.e.\ `below the light cone'. Quantum effects do not give any
contribution to spectral densities here (and more generally below
the threshold for zero-temperature decays), and the purely thermal
contribution is not problematic. Even in HTL approximation, thermal
effects give non zero spectral densities below the light cone: this
describes `Landau damping' i.e.\ the fact that particles  exchange
energy with the thermal plasma. However the HTL approximation cannot
be applied at $k\sim T$ (a region relevant for us, since $g\sim 1$),
and indeed it misses one key physical fact: {\em at $k \gg T$
spectral densities get suppressed by an exponential Boltzmann
factor}. Indeed the  thermally averaged coupling of a particle with
large momentum
 $k\gg T$ is small, since very few of the particles in the plasma have
the large momentum demanded by energy-momentum conservation.
This Boltzmann suppression of the spectral density below the light cone
is the main difference between fig.\fig{Fermionm}a (HTL approximation)
and fig.\fig{Fermionm}b (full one loop),
and makes the gravitino production rate about $50\%$ smaller than
what one would find by applying the HTL approximation at all momenta,
outside its domain of validity $p\ll T$.
%

%Furthermore, going beyond the HTL approximation creates a continuum also
%above the light cone. The rest of  this section discusses the physical interpretation of this result,
%explaining why we will neglect it.
%It just describes a smearing of the particle and plasmon poles:
%as well known from the $T=0$ analogue
%this just means that they acquire a finite width $\Gamma$,
%as illustrated in fig.\fig{Fermionm}.

\subsection{Vector and gaugino propagators}
We are now going to present how spectral densities are practically used.
We need the resummed propagators for the vector with four-momentum $K=(k_0,\mb{k})$
and the gaugino with four-momentum $Q=(q_0,\mb{q})$ in the loop.
We employ non-time ordered propagators (as they allow slightly cleaner formul\ae{} than
imaginary parts of propagators), denoted with a $^<$ in the notation of~\cite{LeBellac} that we follow.
Thermally resummed
propagators are denoted with a $^*$: they are
(see appendices~\ref{thermal},~\ref{thermalF} for more details)
\begin{eqnsystem}{sys:HTLprop<}
^*S^<(Q) &=&\frac{f_F(q_0)}{2}\bigg[(\gamma_0 - \mb{\gamma}\cdot\hat{\mb{q}})
\rho_+(Q)+(\gamma_0 + \mb{\gamma}\cdot\hat{\mb{q}})
\rho_-(Q)\bigg], \\
^*D_{\mu\nu}^<(K) &=& f_B(k_0)\bigg[ \Pi_{\mu\nu}^T \rho_T(K)+\Pi_{\mu\nu}^L \frac{|\mb{k}|^2}{K^2}\rho_L(K)+
\xi \frac{k_\mu k_\nu}{K^4}\bigg].\label{eq:rhoV}
\end{eqnsystem}
Some explanations are in order.
First,  $q_0>0$ or $k_0>0$ describes a fermion or a vector in the final state,
and $q_0<0$ or $k_0<0$ describes a fermion or a vector in the initial state:
this convention allows to compactly describe all possible processes. Indeed
the factors
\begin{eqnsystem}{sys:fBF}
f_B(k_0)&\equiv& \frac{1}{e^{k_0/T}-1} = \left\{\begin{array}{ll}
n_B & \hbox{if $k_0>0$}\\
-(1+n_B) & \hbox{if $k_0<0$}
\end{array}\right.\\
f_F(q_0)&\equiv& \frac{1}{e^{q_0/T}-1}=
\left\{\begin{array}{ll}
n_F & \hbox{if $q_0>0$}\\
1-n_F& \hbox{if $q_0<0$}
\end{array}\right.\end{eqnsystem}
give the usual statistical factors: $-n$ (number of particles in the initial state) or $1\pm n$ (stimulated emission or Pauli-blocking in the final state),
where $n_{B,F}(E)\equiv 1/(e^{|E|/T}\mp 1)$ are the usual Bose-Einstein and Fermi-Dirac distributions.

\medskip

Second,
$\rho_+$, $\rho_-$, $\rho_T$, $\rho_L$ are the spectral densities for
the fermion, fermion pole, transverse vectors and longitudinal vectors respectively.
As discussed in the previous section, we can keep the HTL pole approximation
outside the light cone, so that
\begin{eqnsystem}{sys:rho}
\rho_\pm(Q) &=& 2\pi \bigg[Z_\pm(q)\,\delta(q_0-\omega_\pm(q))+Z_\mp\,
\delta(q_0+\omega_\mp (q))\bigg] + \rho_\pm^{\rm cont}(Q),\\
\rho_{L,T}(K) &=& 2\pi \bigg[Z_{L,T}(k)\,\delta(k_0-\omega_{L,T}(k))-Z_{L,T}\,
\delta(k_0+\omega_{L,T} (k))\bigg] + \rho_{L,T}^{\rm cont}(Q).
\end{eqnsystem}
In HTL approximation the residues at the poles are given in terms of
the pole positions $\omega_\pm(q)$ and $\omega_{L,T}(k)$ as~\cite{WeldonVector, WeldonFermion,LeBellac}
\beq
Z_\pm=\frac{\omega_\pm^2-q^2}{2m_F^2},\qquad
Z_L = \frac{\omega_L(\omega_L-k^2)}{k^2 (k^2+2m_V^2-\omega_L^2)},\qquad
Z_T=\frac{\omega_T(\omega_T^2-k^2)}{2m_V^2 \omega_T^2-(\omega_T^2-k^2)^2}.
\eeq
These formul\ae{} tell that residues for
longitudinal and hole excitations are exponentially suppressed at energies larger than $gT$:
they are low-energy collective phenomena.
The continua $\rho^{\rm cont}$ only
exist below the light cone, at $|q_0|<q$ and $|k_0|<k$.
The spectral densities satisfy sum rules such as
\beq\label{eq:sumrules}
\int_{-\infty}^{+\infty} \frac{dq_0}{2\pi} \rho_\pm(Q)=1,\qquad
\int_{-\infty}^{+\infty} \frac{dk_0}{2\pi} \rho_T(K)=1,\qquad
\int_{-\infty}^{+\infty} \frac{dk_0}{2\pi} \rho_L(K)=\frac{2m_V^2}{3k^2}\eeq
and the continuum turns out to contribute $\sim (10\div 20)\%$ less than
the poles.
Eq.\eq{sumrules} means that the number density of
longitudinal vectors diverges at $k\to 0$,
but this leaves finite gravitino rates thanks to the $d^3k$ integration factor.
In the $T=0$ limit $\omega_\pm(q) =\pm q$, $\omega_{L,T}(k)=k$ and one
can check that the standard expressions for the propagators are recovered.
Notice that $\rho_{L,T}$ have dimensions mass$^{-2}$, while $\rho_\pm$ have dimensions
mass$^{-1}$.

\setcounter{equation}{0}
\section{Gravitino production rate due to decay effects}

\label{decay} We can now compute the imaginary part of diagram D in
fig.\fig{Feyn1}, and extract from it the gravitino production rate.
Using the gravitino $\leftrightarrow$ Goldstino equivalence,
eq.~(\ref{eq:ET}), diagram D is obtained from eq.~(\ref{eq:Pi<}) by
inserting the quadratic parts of the MSSM supercurrent
(\ref{eq:4-comp}) and of its divergence (\ref{eq:dS}):
\begin{align*}
S_{(2)}^{\mu}  &
=-\sum_{N=1}^{3}\frac{1}{4}F_{\nu\rho}^{(N)}[\gamma^{\nu},\gamma^{\rho
}]\gamma^{\mu}\gamma^{5}\lambda^{(N)}-\sqrt{2}\left[
(\partial^{\nu}\phi
_{i})^{\ast}(\gamma^{\nu}\gamma^{\mu}\xi_{L}^{i})+(\partial^{\nu}\phi
_{i})(\gamma^{\nu}\gamma^{\mu}\xi_{R}^{i})\right]  ,\\
(\partial\cdot S)_{(2)}  & =-\sum_{N=1}^{3}\frac{M_N}{4}\Op_N,\qquad
\Op_{N}=F_{\mu\nu}^{(N)}[\gamma_{\mu},\gamma_{\nu}] i\gamma^5 \lambda^{(N)}
\end{align*}
where $N$ runs over the three factors of the MSSM gauge group, and
$F_{\nu \rho}^{(N)}$ here stands for the linearized part of the
corresponding field strength. We ignored soft-breaking squared
masses of scalars, as they have higher dimension than gaugino masses
$M_N$. The contribution to $\Pi^<$ from diagram D is
\begin{align}
\Pi^{<}(P)  & =\frac{1}{4\bar{M}_{\mathrm{Pl}}^{2}}\left[
\mathrm{Tr}\langle{\bar{S}}_{(2)}^{\mu}\Pi^{3/2}_{\mu\nu}S_{(2)}^{\nu}%
\rangle_{T}-\frac{2}{3m_{3/2}^{2}}\mathrm{Tr}\langle(\partial\cdot\bar
{S})_{(2)}\Pl(\partial\cdot S)_{(2)}\rangle_{T}\right]  \\
& =\sum_{N=1}^{3}\frac{1}{32 \bar{M}_{\mathrm{Pl}}^{2}}\left(  1+\frac{M_{N}%
^{2}}{3m_{3/2}^{2}}\right)  \mathrm{Tr}\langle\bar{\Op}_{N}\Pl\Op_{N}%
\rangle_{T}.\label{eq:magic}
\end{align}
We now explain how eq.\eq{magic} is obtained.
The Goldstino part, proportional to $M_N^2/3 m_{3/2}^2$, is straightforward.
We emphasize that
the divergence of the supercurrent is evaluated before evaluating its thermal matrix element.
Indeed, while thermal masses na\"{\i}vely look like SUSY-breaking terms of order
$g^2T$, they actually do not contribute to $\partial_\mu S^\mu$,
and a mistake about this issue would make
the Goldstino rate qualitatively wrong~\cite{wrong,LR,Ellis}
(see also appendix~\ref{gravitino}).
Indeed, despite the nice formalism employed to compute them
(periodic and anti-periodic boundary conditions in
imaginary time for bosons and fermions respectively),
thermal effects just are one particular
background:  no background affects the operator equations of motion, such that
a supercurrent which is conserved at $T=0$ remains conserved at finite $T$.\footnote{Although this is not relevant for us, we can be more
precise: thermal effects spontaneously break supersymmetry in the
visible sector, and the associated thermal Goldstino mode was
identified with a particular collective excitation~\cite{SUSYT}. The
conservation of the supercurrent at finite temperature is therefore
analogous to how electroweak gauge currents remain conserved despite
the Higgs vev.
However, since the thermal Goldstino is a low energy phenomenon,
we don't know how to extend it to write an
explicit conserved  supercurrent that also holds at energies $E\sim T$.}
%As we mentioned above, our computation has a
%residual gravitino gauge dependence which as discussed in appendix A
%is most likely related to the fact that we do not compute
%corrections to the gravitino
%vertices. Therefore we need to enforce $\partial_{\mu}{}S_{\mu}\symbol{126}%
%m_{\text{soft}}$ by hand. The simplest way to do it is to compute the  when
%computing the divergence in absence of softYou can say that we only know
%$S^{\mu}$ and

For the remaining massless gravitino part, we insert the explicit value of the gravitino
polarization tensor (\ref{eq:ripetuta}) and get two terms of the
form
\begin{equation}
\mathrm{Tr}\langle{\bar{S}}_{(2)}^{\mu}\Pl S_{(2)}^{\mu}\,\rangle
_{T}+ \frac{1}{2} \mathrm{Tr}\langle\Ssl_{(2)}\Pl\overline{\Ssl}_{(2)}%
\,\rangle_{T}\label{eq:decompo}%
\end{equation}
where ${}\Ssl_{(2)}\equiv\gamma_{\mu}{}\,S_{(2)}^{\mu}$. We can now
perform simplifications that only employ the known Dirac-matrix
structure of ${S}_{(2)}^{\mu}$:

\begin{itemize}
\item The vector/gaugino contributions obey $\Ssl_{(2)}=0$ (thanks to
$\gamma_{\mu}[\gamma_{\alpha},\gamma_{\beta}]\gamma^{\mu}=0$), such
that only the first term of eq.\eq{decompo} contributes. It is
reduced to the same
operator $\Op_N$ using the $\gamma_{\mu}\gamma_{\alpha}\gamma^{\mu}%
=-2\gamma_{\alpha}$ identity and taking into account that the
thermally corrected gluino propagator has the same $\gamma$-matrix
structure as the massless propagator. This leads to the
$1+M_{N}^{2}/3m_{3/2}^{2}$ prefactor in eq.\eq{magic}.

\item The quark/squark contributions vanish, thanks to a cancellation between
the two terms in eq.\eq{decompo}. Indeed, by applying the $\gamma_{\mu}%
\gamma_{\alpha}\gamma^{\mu}=-2\gamma_{\alpha}$ identity (one time in
the first term, and two times in the second term) both terms reduce
to the matrix
element $\mathrm{Tr}\langle\bar{\xi}_{R}(\ds\varphi)\Pl(\ds\varphi^{\ast}%
)\xi_{L}\rangle$, with opposite coefficients.
\end{itemize}

We don't know if there is some deeper reason dictating these
cancellations such that the full result is controlled by the thermal matrix
element of the operator $\Op_{N}$ times the prefactor
$1+M_{N}^{2}/3m_{3/2}^{2}$.
A general proof of this result would
allow to get the full production rate from the simple Goltstino rate
according to eq.\eq{magic}.

%A similar phenomenon was previously encountered
%(and partly explained with arguments similar to ours)
%in~\cite{Buch}, where it was noticed the differential cross sections for the production
%of the spin-1/2 and spin-3/2 gravitino components are proportional, process by process, with the same
%prefactor. A similar relation is encountered in our computation of the production rate due to the
%top Yukawa coupling (section~\ref{top}).

\bigskip

For completeness we mention that we have
studied thermal corrections to the $\Op_N$ operators in HTL
approximation. As well known gauge vertices $g$ receive very large
thermal corrections of order $g(1+g^{2}T^{2}/k^{2})$ where $k\ll T$
(HTL approximation) is some external momentum: their presence would
be problematic, as they seem to describe infra-red divergent effects
(see e.g.\ section 10.3 of~\cite{LeBellac}). In the case of gauge
vertices these corrections are demanded by gauge invariance:
different diagrams combine such that $\Lag_{\rm HTL}$ of eq.\eq{LHTL}
contains the gauge-covariant derivative $D$.
On the contrary Yukawa couplings do not receive these problematic HTL
corrections. We verified that the Goldstino vertex $\chi\Op_N$ does
not receive any HTL correction.\footnote{The basic reason is the
following. Since the Goldstino vertex has dimension 5, by
dimensional analysis it receives gauge
corrections of order $g^{2}\int d^{4}K(K^{3})_{\mu}/[K^{2}(K+P_{1}%
)^{2}(K+P_{2})^{2}]$ where $(K^{3})_{\mu}$ denotes any vector formed
with 3 powers of $K$: it necessarily contains the combination
$K^{2}$, that, as explained in~\cite{LeBellac}, does not lead to HTL
vertices.} Beyond the HTL limit there will be corrections suppressed
by powers of $g/\pi$, that we can ignore.

\subsection{Gravitino propagator}
We now restart from eq.\eq{magic} and
explicitly compute the imaginary part of the gravitino propagator
with four-momentum $P=(p_0,\mb{p})=K+Q$,
summed over its polarizations:
% PL...PR was replaced by    i gamma5...i gamma5    *  1/2
\beq
\Pi^<(E) =\sum_{N=1}^3\bigg(1+\frac{M_N^2}{3m_{3/2}^2}\bigg) \frac{n_N}{16(2\pi)^2 \bar{M}_{\rm Pl}^2 } \int \frac{d^4K}{(2\pi)^4} {\rm Tr}
[\Pl [\Ksl,\gamma_\mu] i\gamma_5 {} ^*{\!}S^<(Q)i\gamma_5 [-\Ksl,\gamma_\nu]
^*D_{\mu\nu}(K)].
\eeq
where $N=\{1,2,3\}$ runs over the three factors of the SM gauge group with $n_N=\{1,3,8\}$
vectors;
$M_N$ are the gaugino masses at zero temperature (renormalized at some scale around $T$).
Inserting the explicit parameterization
$$ K=(k_0, k,0,0),\qquad
Q=(q_0,q\cos\theta_q,q\sin\theta_q),\qquad
P=(p,p \cos\theta_p,p\sin\theta_p)$$
for the vector, gaugino and gravitino four-momenta respectively one finds
\begin{eqnarray}
\Pi^<(p) &=& \sum_{N=1}^3 p\bigg(1+\frac{M_N^2}{3m_{3/2}^2}\bigg) \frac{n_N}{ \bar{M}_{\rm Pl}^2 } \int \frac{d^4K}{(2\pi)^4} \nonumber
f_B(k_0) f_F(q_0)\times\\
&& \bigg[\rho_L(K)\rho_-(Q)  k^2 \cos^2 \frac{\theta_p+\theta_q}{2}+\nonumber
\rho_L(K)\rho_+(Q)  k^2 \sin^2 \frac{\theta_p+\theta_q}{2}+\\
&&+\rho_T(K)\rho_+(Q) \bigg((k^2+k_0^2)(1+\cos\theta_p \cos\theta_q)-2k k_0 (\cos\theta_p + \cos\theta_q)\bigg)+\\
\nonumber&&+\rho_T(K)\rho_-(Q) \bigg((k^2+k_0^2)(1-\cos\theta_p \cos\theta_q)-2k k_0 (\cos\theta_p- \cos\theta_q)\bigg)\bigg]
\end{eqnarray}
To compute the total rate $\gamma_D$ using eq.\eq{ImProp}
%\beq \gamma_D = \frac{dN}{dV~dt}= \int d\vec{P} ~\Pi^<(P),\qquad
%d\vec{P}\equiv  \frac{d^3p}{2p_0(2\pi)^3}\eeq
it is convenient to multiply by $1 = \int d^4 Q~\delta^4(K-P-Q)$,\footnote{This step also allows
to see that the seemingly esoteric expression is actually
equivalent to what one would na\"{\i}vely guess from
the kinetic theory, if spectral densities are treated like parton densities
$$\frac{dN}{dV\,dt\, d\vec{P}}=  \sum \int_{q_0,k_0\ge 0} \frac{d^4 Q}{(2\pi)^4}\frac{d^4 K}{(2\pi)^4}
\rho_\pm(Q) \rho_{L,T}(K) |\Amp|^2 (2\pi)^4 \delta^{4}(P\pm Q \pm K)
\cdot\hbox{(statistical factors})$$ where the sum is over all
polarizations, gauge indices, gravitino production processes with
amplitudes $\Amp$. As discussed around eq.~(\ref{sys:fBF}), the
factors $f_B(k_0)$ and $f_F(q_0)$ reproduce the usual statistical
factors, $1\pm n$ or $-n$.} perform the non-trivial angular
integrations over $\theta_p$ and $\theta_q$, obtaining
%\begin{eqnarray}
%\gamma_D &=&\nonumber
%\frac{1}{2(2\pi)^5} \sum_{N=1}^3\bigg(1+\frac{M_N^2}{3m_{3/2}^2}\bigg) \frac{n_N}{\bar{M}_{\rm Pl}^2 } \int_0^\infty dq~dk \int_{-\infty}^{+\infty}dq_0~dk_0~k f_B(k_0) f_F(q_0) \times\\  \nonumber
%&&  \times \bigg[
%\rho_L(K)\rho_-(Q)  (p-q)^2 [(p+q)^2-k^2]+
%\rho_L(K)\rho_+(Q)  (p+q)^2[k^2-(p-q)^2]+\\
%&&+\rho_T(K)\rho_+(Q) ((p+q)^2-k^2)\bigg((1+k_0^2/k^2)(k^2+(p-q)^2)-4 k_0(p-q)\bigg)+\\
%\nonumber
%&&+\rho_T(K)\rho_-(Q) (k^2-(p-q)^2)\bigg((1+k_0^2/k^2)(k^2+(p+q)^2)-4 k_0(p+q)\bigg)\bigg].
%\end{eqnarray}
\beq\label{eq:res}
 \gamma_D =\frac{T^6}{2(2\pi)^3\bar{M}_{\rm Pl}^2}  \sum_{N=1}^3 n_N \bigg (1+\frac{   M_N^2}{3  m_{3/2}^2}\bigg)
 f_N,
\eeq
where
\begin{eqnarray}
f_N&=&T^{-6}\int_{-\infty}^{+\infty}dq_0~dk_0~k f_B(k_0) f_F(q_0)
\times\nonumber\\
\nonumber &&  \times \bigg[ \rho_L(K)\rho_-(Q)  (p-q)^2
[(p+q)^2-k^2]+
\rho_L(K)\rho_+(Q)  (p+q)^2[k^2-(p-q)^2]+\\
&&+\rho_T(K)\rho_+(Q) ((p+q)^2-k^2)\bigg((1+k_0^2/k^2)(k^2+(p-q)^2)-4 k_0(p-q)\bigg)+\label{eq:f_N}\\
\nonumber &&+\rho_T(K)\rho_-(Q)
(k^2-(p-q)^2)\bigg((1+k_0^2/k^2)(k^2+(p+q)^2)-4
k_0(p+q)\bigg)\bigg].
\end{eqnarray}
The dimensionless coefficients $f_N$ are positive:
each term in the square brackets is positive in the allowed region,
except $\rho_T$ that becomes positive after being multiplied by
$f_B(k_0)$. The integration range is restricted by momentum
conservation, $\mb{p}+\mb{k}+\mb{q}=0$, i.e.\  $|k-q|\le p=k_0+q_0
\le k+q$: any side of a triangle cannot be longer than the sum of
the other two or shorter than their difference.

\medskip

The last two equations generalize eq.\ (38) of~\cite{Buch}, who considered the vector/gaugino loop in the
limit of hard gravitino and soft vector (small $k_0\ll p,T$, such
that $f_B(k_0)\simeq T/k_0$) and neglected the gaugino thermal mass
(i.e.\ $\rho_- \simeq 0$ and $\rho_+\simeq 2\pi \delta(q_0 - q)$
such that $q=q_0 = p-k_0$).

%We approximate the spectral densities outside the light cone as $\delta$-function poles
%(using full expressions they would be narrow bells, making numerical integration
%difficult for our limited computing power), such that we have four types of contributions:
%pole-pole, continuum-continuum, (vector pole)-(gaugino continuum) and (vector continuum)-(gaugino pole). A vector can be either longitudinal or transverse, in the initial state or in the final state,
%and similarly for the gaugino.
%

\subsection{Decay contribution to the gravitino production rate}\label{fNsummary}
%After summing all decay contributions we get the decay contribution to the gravitino production
%rate per space-time volume as
%\beq\label{eq:res}
% \gamma_D =\frac{T^6}{2(2\pi)^3\bar{M}_{\rm Pl}^2}  \sum_{N=1}^3 n_N \bigg (1+\frac{   M_N^2}{3  m_{3/2}^2}\bigg) f_N(\frac{m_{V_N}}{T})
%\eeq
In conclusion, the decay contribution to the gravitino production rate
per space-time volume is given by eq.\eq{res}. The
coefficients $f_N$ have to be evaluated numerically. We approximate
the spectral densities outside the light cone as $\delta$-function
poles (using full expressions they would be narrow bells, making
numerical integration difficult for our limited computing power),
such that we have four types of contributions: pole-pole,
continuum-continuum, (vector pole)-(gaugino continuum) and (vector
continuum)-(gaugino pole). A vector can be either longitudinal or
transverse, in the initial state or in the final state, and
similarly for the gaugino.

The resulting coefficients $f_N$ depend on the gauge couplings and on
the content of matter charged under the given gauge group; 
in the MSSM it is convenient to parametrize them as functions of the thermal vector
masses $m_{V_N}$ listed in eq.\eq{mVn}: \beq f_N\equiv f_N(\frac
{m_{V_N}}{T}) .\eeq For example $m_{3}\approx 1.3T$ for the gluon at
$T\approx 10^{9}\GeV$. 
The functions $f_N$ are plotted in
fig.\fig{res}. In HTL approximation there would be a unique
$N$-independent function $f$, and the functions $f_N$ turn out to be
somewhat different, depending on the relative amount of vector and
chiral multiplets present within each group. In  non-minimal models
with more chiral multiplets than in the MSSM, one would have to add
their extra contributions to vector thermal masses, and to slightly
revise the functions $f_N$.

Finally, let us try to discuss the accuracy of our result.
Thermal corrections to the pressure have been computed up to high
orders in $g_3$~\cite{largeNf}:
these computations can be used to see how convergent the perturbative
expansion is in practice.
In the favorable limit $N_f\gg N_c$ (where $N_f$ is the number of
flavors and $N_c$ is the number of colors)
the perturbative expansion for the pressure remains accurate up to $m_
{V}/T \approx 1$
%  i.e. g_eff^2 = 6 in his notation
(fig.~1 of~\cite{largeNf}).
This presumably also applies to our case, as
SUSY-QCD has a set of fermions and scalars that give the same
contribution
to the gluon thermal mass as $N_f = 21$ flavors in the fundamental.

Furthermore, AdS/CFT techniques should allow to compute the large
coupling limit of the gravitino emission rate in some supersymmetric theory, maybe not unrealistically
different from SUSY-QCD. This could be done analogously to how \cite{Starinets} used AdS/CFT to compute the photon emission rate from strongly coupled $\mathcal{N}=4$ SYM in the large $N_c$ limit.
By analogy, we expect that at strong coupling the gravitino rate functions $f_N$ will have a finite limit, $N_c$-independent up to $1/N_c$ corrections.
% i.e. \gamma=const. T^6 where const=N^2 up to 1/N corrections

\begin{figure}[t]
\begin{center}\vspace{-1cm}
$$\includegraphics[width=0.7\textwidth]{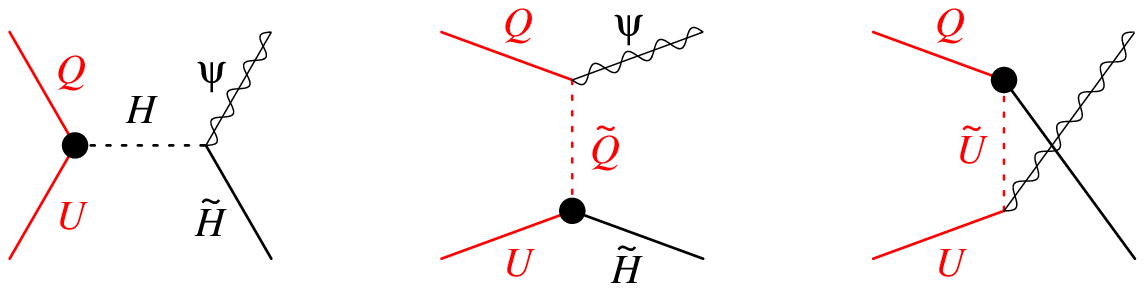}$$
\caption{\label{fig:top4}\em {\bf Top scatterings}.
Feynman diagrams contributing to $QU \to \Psi_\mu \tilde{H}$.
The $\bullet$ denotes the top Yukawa coupling.}
$$\includegraphics[width=\textwidth]{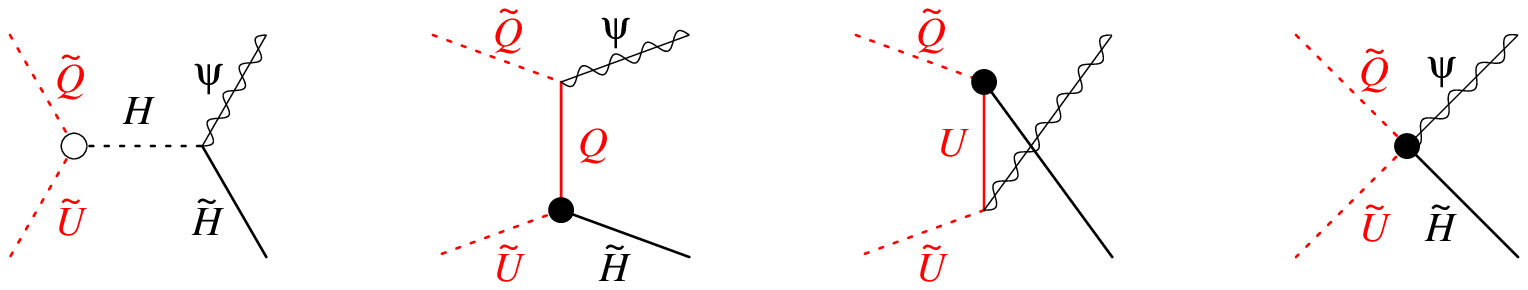}$$
\caption{\label{fig:top22}\em {\bf Top scatterings}.
Feynman diagrams contributing to $\tilde{Q}\tilde{U} \to \Psi_\mu \tilde{H}$.
The $\bullet$ denotes a coupling proportional to $\lambda_t$, and
$\circ$ its $A$-term.}
\end{center}
\end{figure}

\setcounter{equation}{0}
\section{Production of gravitinos due to the top Yukawa}\label{top}
Previous works considered gravitino production due to the
$g_3$, $g_2$ and $g_Y$ gauge couplings;
the top quark Yukawa, $\lambda_t ~ QUH$,
also has a sizable coupling $\lambda_t$.
There are two main kind of scattering processes:
\begin{itemize}
\item[a)]  Scatterings involving fermions only,
such as $QU \to \Psi \tilde{H}$: fig.\fig{top4} shows the relevant
Feynman diagrams. They would populate only the spin 3/2 component of
the gravitino, as only dimension-2 soft terms enter these diagrams,
so that Goldstinos are not produced. However, an explicit
computation shows that the dominant contribution, of order
$T^6/M_{\rm Pl}^2$, vanishes.

\item[b)] Scatterings involving two fermions and two scalars, such as
$\tilde{Q}\tilde{U}\to \Psi \tilde{H}$: fig.\fig{top22} shows the
relevant Feynman diagrams. The first diagram involves $A_t$, the
dimension-1 $A$-term of the top Yukawa coupling, and populates the
spin 1/2 component of the gravitino. The other three diagrams
populate the spin 3/2 component.
\end{itemize}
The total result is:
\beq \label{eq:sigmatop}
\sum_{\rm all} | \Amp(\hbox{top scatterings})|^2 = 72 \frac{\lambda_t^2}{\bar M_{\rm Pl}^2}
(1 + \frac{A_t^2}{3m_{3/2}^2}) s
\eeq
where $s=(P_1+P_2)^2$ is the usual kinematical variable.
The corresponding gravitino production rate is
\beq\label{eq:gammatop}
\gamma_{\rm top} =1.30  \frac{9\lambda_t^2 T^6}{2\bar{M}_{\rm Pl}^2\pi^5}(1 + \frac{A_t^2}{3m_{3/2}^2})
\eeq
where the numerical factor $1.30$ is the correction due to the Fermi-Dirac and Bose-Einstein
factors with respect to the Boltzmann approximation.

In the language of previous sections, eq.\eq{gammatop} is the scattering contribution.
We now explain why it also is our total result.
First, it happens to be infra-red convergent: the potentially divergent contributions
given by the modulus squared of the $t$-channel and $u$-channel diagrams in fig.\fig{top22}
actually vanish.
Therefore, unlike in the case of gauge scatterings,
the inclusion of thermal masses is not necessary for obtaining a finite result.
Furthermore, including thermal effects along the lines of the previous sections does not affect
the final result.
Indeed the top Yukawa coupling gives a thermal mass for top, stops (and higgs and higgsinos):
the resulting quark/squark/gravitino (and higgs/higgsino/gravitino) decay rates have been
computed in section~\ref{decay} for generic thermal masses, and vanish.
Consistency requires that the subtracted top scattering rate
equals the total scattering rate of eq.\eq{gammatop}, and indeed the
subtracted terms are
the modulus squared of the $t$-channel and $u$-channel diagrams in fig.\fig{top22},
which vanish.

\medskip

Again, all these cancellations have a simple interpretation:
they are the ones needed such that the production rate for
the spin 3/2 components of the gravitino equals the production rate for
the spin 1/2 Goldstino components, up to the prefactor in eq.\eq{sigmatop}.
Indeed, using the gravitino/Goldstino equivalence,
the Goldstino production rate can be equivalently computed from one single diagram
that only involves the single {\em quartic} Goldstino coupling,
$$ A_t \lambda_t ~\hbox{Goldstino}~(\hbox{higgsino} ~\hbox{squark}~\hbox{squark}^* +
\hbox{quark}~\hbox{squark}^*~\hbox{higgs} + \hbox{h.c.}),$$
such that decay contributions and subtracted scattering rates simply do not exist for the Goldstino.

% FROM SLAVA:
%The true result is equal to the Boltzmann one times the factor (7.9+8.3)/(6.2*2)=1.30
%Here 7.9 is the scalar+scalar->gravitino+fermion part, and 8.3 is the scalar+fermion->gravitino+scalar
%In the Boltzmann approximation both are equal 6.2 (in arbitrary units).

%
%This is comparable to the (subtracted) SU(3) effect, which was not very important.
%To
%keep the MSSM alive one sometimes wants to get the maximal Higgs mass, obtained for
%a somewhat large $A_t\approx \sqrt{6}m_{\tilde{t}}$ at the weak scale.

\begin{figure}[t]
$$\includegraphics[width=10cm]{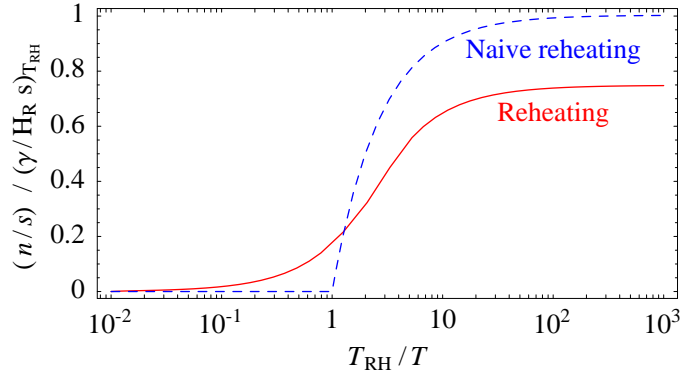}$$
\caption{\label{fig:TRH}\em Evolution of the gravitino abundancy
$n/s$ in units of $\gamma/H_Rs$ at $T=T_{\rm RH}$
for na\"{\i}ve instant reheating and for the conventional model of reheating.}
\end{figure}

\setcounter{equation}{0}
\section{Boltzmann equations with reheating}\label{reh}
We here compute the gravitino abundance by integrating the relevant
Boltzmann equations. While previous works ignored the
history of the universe prior to its reheating
(from the point of view of computing gravitino production this in practice
amounts to assuming that the Big Bang started at the
maximal temperature $T_{\rm RH}$),
%assumed the big-bang
%suddenly started at the maximal temperature $T_{\rm RH}$,
we here
follow the standard definition of the reheating temperature $T_{\rm
RH}$, where MSSM particles are progressively reheated by the energy
released by some non-relativistic energy density $\rho_\phi$, which
could describe e.g.\ an oscillating inflaton field, or some
non-relativistic particle decaying into MSSM particles.\footnote{
Alternatively, some of the flat directions present in the MSSM supersymmetric potential might develop large vevs during inflation.
There is a debate in the literature whether such condensates can be sufficiently long-lived to affect reheating~\cite{flat}.
For simplicity, we here do not consider these possible but model-depenent phenomena.}
In both cases the relevant Boltzmann equations are
\beq\label{eq:Boltzt}\left\{\begin{array}{l} \dot\rho_\phi +
3H\rho_\phi = -\Gamma_\phi \rho_\phi \cr \dot\rho_R + 4 H \rho_R =
\Gamma_\phi \rho_\phi \cr \dot n_{3/2} + 3 H n_{3/2} = \gamma
\end{array}\right.\eeq
where $n_{3/2}$ is the gravitino number density summed over its polarizations,
a dot denotes $d/dt$,
$H = \dot R/R = \sqrt{8\pi(\rho_\phi  + \rho_R)/3}/M_{\rm Pl}$ is the expansion factor,
$\rho_R=\pi^2 g_*  T^4/30$ is the energy density of MSSM radiation at temperature $T$
(with $g_*=228.75$, up to ${\cal O}(g^2)$ corrections, and up to adding right-handed neutrinos), and
$\Gamma_\phi$ parameterizes the decay width of $\rho_\phi$.
The reheating temperature $T_{\rm RH}$ is defined in terms of $\Gamma_\phi$ as the
temperature at which~\cite{book}
\beq
\Gamma_\phi = H_R\equiv \frac{1}{M_{\rm Pl}}\sqrt{\frac{8\pi}{3}\rho_R(T_{\rm RH})}
\qquad\hbox{i.e.}\qquad
T_{\rm RH}= \left[ \frac{45}{4\pi^3 g_*}\,\Gamma_\phi^2
M_{\rm Pl}^2\right]^{1/4}  .\eeq
It is convenient to rewrite equations\eq{Boltzt} in terms of $Y(z)$,
where $z\equiv T_{\rm RH}/T$ and
$Y\equiv n_{3/2}/s$, with $s=4\rho_R/3T$ being the MSSM entropy density.
Following~\cite{leptog} one gets
\begin{equation}\label{eq:dY/dz}
\left\{\begin{array}{rcl}\displaystyle
  HZz\frac{d\rho_\phi}{dz}&=& -
{3H\rho_\phi}-{\Gamma_\phi\rho_\phi}\,   ,\\[4mm]
\displaystyle
sHZz \frac{dY}{dz} &=& 3sH(Z-1)Y+\gamma
\end{array}\right.\eeq
where \beq
Z =-\frac{\dot\rho_R}{4\rho_R H}= 1 - \frac{\Gamma_\phi \rho_\phi}{4 H \rho_R}.
\end{equation}
To clarify the physical meaning of $T_{\rm RH}$, we emphasize that
$T_{\rm RH}$ is not the maximal temperature; however what happens at
$T\gg T_{\rm RH}$ gets diluted by the entropy release described by
the $Z-1$ factor ($Z\simeq 3/8$ at $T\gg T_{\rm RH}$ and $Z\simeq 1$
at $T\ll T_{\rm RH}$). In our case $\gamma(T)\propto T^6$ and the
solution is \beq \label{eq:Yres} Y(T\ll T_{\rm RH}) = 2\left.
\frac{\gamma}{Hs}\right|_{T=T_{\rm RH}}=0.745 \left.
\frac{\gamma}{H_Rs}\right|_{T=T_{\rm RH}} = 6.11~10^{-12}
\frac{T_{\rm RH}}{10^{10}\GeV} \frac{\gamma|_{T=T_{\rm RH}}}{T_{\rm
RH}^6/\bar M_{\rm Pl}^2}. \eeq Fig.\fig{rate} shows our results for
the dimensionless order-one combination $\gamma/(T^6/\bar{M}_{\rm
Pl}^2)$ that appears in~\eq{Yres}. Notice that the gravitino
abundance is proportional to it, and to $T_{\rm RH}$: the large
power $\gamma(T)\propto T^6$ gets almost compensated by cosmological
factors.

In previous analyses $\rho_\phi$ was ignored and the 
`instantaneous reheating' approximation was used, which amounts to
start the Big Bang from a maximal temperature 
$T=T_{\rm RH}$: this gives a slightly larger gravitino abundance
\beq Y(T\ll T_{\rm RH}) = \left.
\frac{\gamma}{H_Rs}\right|_{T=T_{\rm RH}} .\eeq Fig.\fig{TRH}
illustrates the different evolution of $Y$ (in arbitrary units)
between the two cases.

\medskip

\begin{figure}[t]
\parbox{0.5\textwidth}{\includegraphics[width=0.5\textwidth]{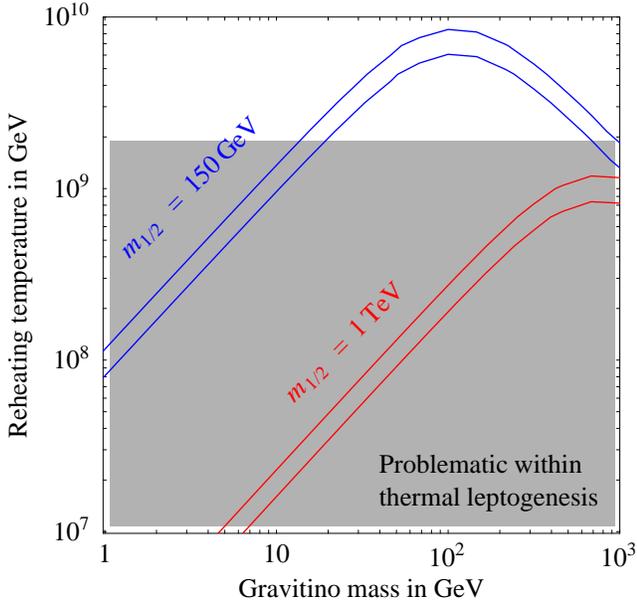}}
\hspace{0.05\textwidth}
\parbox{0.45\textwidth}{
\caption{\label{fig:DM}\em The bands show the region where the
thermal gravitino abundance equals the DM abundance ($3\sigma$
regions), assuming unified gaugino masses with $m_{1/2}=150\GeV$
(roughly the minimal value allowed by present data) or
$m_{1/2}=1\TeV$ at the unification scale, and negligible $A_t$.
Model-dependent issues are here ignored, including who is the LSP and the NLSP.
Successful thermal leptogenesis with zero initial right-handed
neutrino abundance is not possible within the gray
band~\cite{DI,leptog}. }}
\end{figure}

\begin{figure}[t]
$$\includegraphics[width=0.75\textwidth]{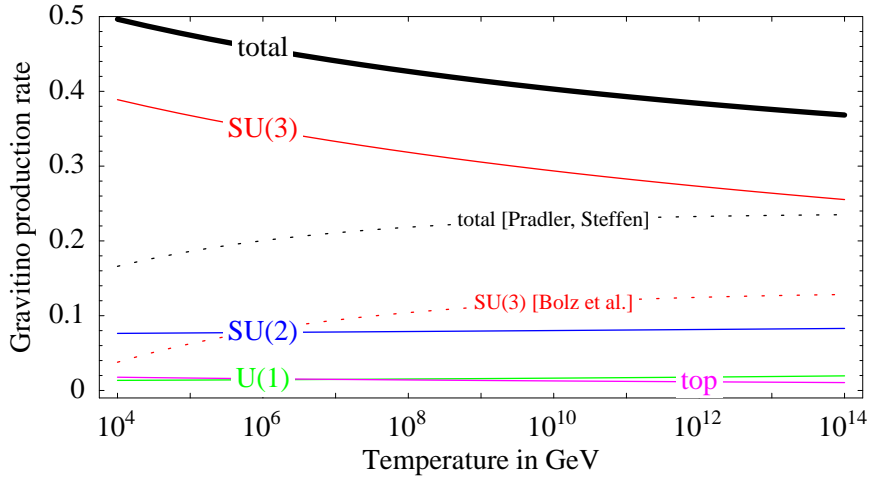}$$
\caption{\label{fig:rate}\em Production rate $\gamma=dN/dV\,dt$
in units of $T^6/\bar M_{\rm Pl}^2$
for the spin-3/2 gravitino components
in the MSSM. The upper curve is the total rate,
and the other continuous curves show the contributions from
$\{g_3,g_2,g_Y,\lambda_t\}$ interactions (summed over decay and scattering processes).
The production rate for the Goldstino spin-1/2 components is obtained by multiplying
these four contributions times $\{M_3^2,M_2^2,M_1^2,A_t^2\}/3m_{3/2}^2$ respectively.
The dotted curve show previous results from~\cite{Buch} and~\cite{postBuch}.}
\end{figure}

Within the standard $\Lambda$CDM cosmological model, present data demand a DM energy density
 $\Omega_{\rm DM} h^2=0.110\pm 0.006$; if DM are non relativistic particles with mass $M\gg {\rm keV}$
 this corresponds to $Y_{\rm DM} = (0.40\pm0.02)\eV/M$~\cite{WMAP}.
 One has $Y=Y_{\rm DM}$ if gravitinos are the observed DM.
Equivalently, one can compute the present gravitino mass density in
terms of their relative entropy $Y$ as \beq \Omega_{3/2} h^2 =
\frac{m_{3/2} Y s_{0}}{\rho_{\rm cr}/h^2}=0.274~10^9~ Y
\frac{m_{3/2}}{\GeV} = 0.00167 \frac{m_{3/2}}{\GeV}\frac{T_{\rm
RH}}{10^{10}\GeV} \frac{\gamma|_{T=T_{\rm RH}}}{T_{\rm RH}^6/\bar
M_{\rm Pl}^2}. \eeq where $\rho_{\rm cr} = 3H_0^2 M_{\rm Pl}^2/8\pi
$ is the critical energy density, $H_0 = 100\, h
\,\hbox{km/sec$\cdot$Mpc}$ is the Hubble constant, the present
entropy density is $s_{\rm 0}  =2\pi^2 g_{*s}T_0^3/45$ with $g_{*s}
= 43/11$ and $T_0=2.725\,{\rm K}$. Fig.\fig{DM} compares the regions
where the thermal gravitino abundance equals the DM abundance with
the regions compatible with standard thermal leptogenesis~\cite{DI}, as
computed in~\cite{leptog} with the same definition of the reheating
process.
This plot ignores all model dependent issues, including who is the LSP and the NLSP.
Let us briefly summarize these
issues~\cite{gravitinoCosmo,Moroi,Viel,altri}.
\begin{itemize}
\item If the gravitino is the stable LSP then
\begin{itemize}
\item if $m_{3/2}\gg {\rm keV}$ the gravitino behaves as cold dark matter and its
energy density  can be at most equal to the DM density.

\item a somewhat stronger constraint applies if the gravitino is lighter,
$T_0 \ll m_{3/2}\circa{<} {\rm keV}$, and consequently behaves as
warm dark matter or radiation.
The Goldstino component of such a light gravitino thermalizes
(unless $T_{\rm RH}$ is as low as possible),
so that this scenario is severely constrained:
assuming that the  Goldstino  was in thermal equilibrium
when $g_* \sim 100$, present data demand
$m_{3/2}\circa{<}16\eV$~\cite{Viel}.

\end{itemize}
An additional contribution to the gravitino energy density,
$\Omega_{3/2}^{\rm extra}\simeq \Omega_{\rm NLSP}m_{3/2}/m_{\rm NLSP}$
(having neglected entropy production, which typically is an excellent approximation)
is  generated by NLSP decays, with mass $m_{\rm NLSP}$ and mass density
$\Omega_{\rm NLSP}$ after their freeze-out.
Weak scale sparticles give $\Omega_{\rm NLSP}\sim 1$, such that this extra contribution is significant
if $m_{3/2}$ is not much smaller than $m_{\rm NLSP}$.

\item If heavier than the LSP (which possibly has mass $m_{\rm LSP}\sim 100 \GeV$),
the gravitino gravitationally decays into the LSP and some SM particles:
\begin{itemize}
\item If $m_{3/2}\gg 10 \TeV$ the gravitino decays before BBN,
generating a contribution to the LSP energy density,
$\Omega_{\rm LSP}\simeq m_{\rm LSP}  \Omega_{3/2}/m_{3/2} $~\cite{gravitinoCosmo} (we neglected the entropy in gravitinos).

\item A lighter gravitino
decays during or after BBN, damaging nucleosynthesis. The resulting
bound on $\Omega_{3/2}$ depends on which SM particles are produced
by gravitino decays, and typically is some orders of magnitude
stronger than the DM bound $\Omega_{3/2} h^2
\circa{<}0.1$~\cite{Moroi}. Very late gravitino decay into photons
would also distort the CMB energy spectrum.
\end{itemize}
\end{itemize}
We recall that we computed thermal production of gravitinos  from MSSM particles
at temperatures $T_{\rm RH}\gg m_{3/2}, m_{\rm soft}$.
The true physics might be different.
For example, the messenger fields with mass $M_{\rm GM}$
employed by gauge mediation models
might be so light that they get thermalized (together with the hidden sector)
and later decay back to MSSM particles,
leaving a thermalized Goldstino.
Eq.\eq{dY/dz} shows that this phenomenon is dominant if $M_{\rm GM}\circa{<}(10\div100) T_{\rm RH}$,
as the gravitino abundance gets washed-out as $Y\propto T^{5}$
during reheating at $T\circa{>}T_{\rm RH}$.

\setcounter{equation}{0}
\section{Conclusions}\label{conclusions}
Previous computations of the thermal gravitino production rate~\cite{Buch,postBuch}
were performed at leading order in small gauge couplings,
finding a rate of the form $\gamma \propto g^2 \ln 1/g$, which
 behaves unphysically when extrapolated to the true MSSM  values of the
 gauge couplings, $g\sim 1$  (see fig.\fig{res}).
 We improved on these results in the following ways:
\begin{enumerate}
\item  We included gravitino production via gluon $\to$ gluino + gravitino and other decays:
these effects first arise at higher order in $g$ (the phase space is
opened by thermal masses), but are enhanced with respect to
scattering processes by a phase-space $\pi^2$ factor, typical of
3-body vs 4-body rates. The gravitino production rate becomes about
twice larger, or more if $M_3 \circa{>} M_{1,2}\gg m_{3/2}$.
\item   We added production processes induced by the top quark Yukawa coupling.
This enhances the gravitino production rate by almost $10\%$
% $3\% $ with respect to our total
or more if $A_t$ is bigger than gaugino masses.

 \item   Finally, we computed the gravitino abundance replacing the instant reheating approximation
 with  the standard definition of the reheating process, where $T_{\rm RH}$ is not
the maximal temperature but defines the temperature
at which inflaton decay ends, ceasing to release entropy.
 This improvement decreases the gravitino abundance by $25\%$ and allows a precise
 comparison with  leptogenesis~\cite{DI}, where reheating was included in~\cite{leptog}.
 \end{enumerate}
 Our result for the gravitino production rate is
 \beq \gamma = \gamma_D + \gamma_{S}^{\rm sub} + \gamma_{\rm top}.\eeq
 where the decay rate $\gamma_D$ (which dominates the total rate)
 is given in eq.\eq{res}, the subtracted scattering rate $\gamma_S^{\rm sub}$ in eq.\eq{Ssub}, and the rate induced by the top Yukawa coupling in eq.\eq{gammatop}.
Fig.\fig{rate} summarizes our results, showing the value of the
dimensionless combination
 $\gamma/(T^6/\bar{M}_{\rm Pl}^2)$ (as well as the values of the single gauge and top contributions to it)
which determines the gravitino abundance as in~\eq{Yres}. In this
plot we assumed $\lambda_t = 0.7$ and a unified $\alpha =1/24$,
renormalized at the scale $M_{\rm GUT}=2\cdot 10^{16}\GeV$.

\medskip

Accessory results scattered through the paper include: a clean
precise re-derivation of gravitino couplings; expressions for
thermal masses in a generic supersymmetric theory; the observation
that they respect supersymmetry at energy much larger than the
temperature; a collection of formul\ae{} for thermal corrections to
vectors (including correct imaginary parts) beyond the Hard Thermal
Loop (HTL) approximation; a possibly non-standard discussion of the
physical meaning of negative spectral densities; a technique that
allows to deal with Coulomb-like infra-red divergences without
introducing an arbitrary splitting scale $k_*$ that satisfies the
problematic condition $gT \ll k_* \ll T$.

\medskip

A curious result simplified our computation:
the differential production rates for the spin 3/2 and for the spin 1/2 (Goldstino)
gravitino component are equal up to a universal prefactor,
despite that
Goldstino couplings (to dimension-1 SUSY-breaking soft terms in the supercurrent)
apparently are much simpler than
the spin-3/2 couplings (gravitational, to the supersymmetric supercurrent).
This equality holds thanks to various cancellations,
such that various troubling contributions automatically drop out from our  computation.

\small\bigskip

\paragraph{Acknowledgements}
We thank R. Barbieri, C. Scrucca, G. Giudice, A. Notari, G. Moore.
A.S.\ thanks R. Rattazzi who suggested that we look into the
higher-order corrections to the gravitino production rate because
this problem involves conceptual issues of academic interest. On the
contrary the 100\% enhancement with respect to previous results is
relevant for phenomenology, and we could compute it without
understanding why we don't need to understand how to write the
supercurrent at finite temperature.

\bigskip\bigskip% \newpage

%
%It would be interesting to sort out the general reason behind
%the cancellations mentioned above,
%but  R. Rattazzi told us that we addressed the issue of higher-order corrections to the
%gravitino production rate because its conceptual issues have academical interest.
%On the contrary the 100\%
%enhancement with respect to previous results is relevant for
%phenomenology, and we could compute it without understanding
%why we don't need to understand how to write the supercurrent at finite temperature.
%We thank R. Barbieri, C. Scrucca, G. Giudice, A. Notari, G. Moore, and R. Rattazzi.

\appendix

\setcounter{equation}{0}
\section{Gravitino propagator and couplings}

\label{gravitino} We here derive the needed gravitino propagator and
couplings, both generic and specialized for the MSSM. The results contained in
this section are not new, but they are useful for two reasons. First, because
all factors ($i,\gamma_{5},P_{L}$, etc) and subtleties should be right, and
are relevant for satisfying the consistency checks that we performed in our
subsequent computations. Second, because we recomputed relevant gravitino
properties in a way that we consider simpler than in previous literature: we
proceed directly, without using Noether and supergravity techniques, which are
unnecessarily cumbersome for our purposes.
We use the standard Weyl spinor and $\gamma$-matrix conventions
corresponding to the signature $(+---)$, see e.g. \cite{Lykken}. The
phase of the gauginos is chosen such that gaugino couplings to
matter are real. We assume a Minkowski background i.e.\ we neglect
the small cosmological constant.

\subsection{The gravitino Lagrangian}

The gravitino is the gauge field associated with local supersymmetry, and
becomes massive by means of a super-Higgs mechanism: `eating' the massless
Goldstino fermion arising when global supersymmetry is spontaneously broken.
This is analogous to a gauge vector that becomes massive via the usual Higgs
mechanism, so that we start recalling some general properties of this well
known simpler case, and this analogy will later allow us to derive gravitino
properties following the same logic.

\subsubsection*{Paradigmatic digression}

%\subsection{Goldstone and Higgs mechanisms}
We thus consider a U(1) gauge symmetry broken by the vev of a charged scalar field
$H$.
%It is convenient to split discussion into two parts.
In the limit of vanishing gauge coupling, a massless Goldstone $\chi$ appears
in the expansion of $H$ around the minimum, $H=v+i\chi$, and $\chi$
transforms (under a U(1) rotation with infinitesimal angle $\varepsilon$) as
$\delta\chi=v\varepsilon$. The total U(1) current is given by
\begin{equation}
J_{\mu}=J_{\mu}^{\text{mat}}-v\, \partial_{\mu}\chi, \label{eq:totalJ}%
\end{equation}
where $J_{\mu}^{\text{mat}}$ is the U(1) current of the rest of the theory
(e.g.\ fermionic matter).
%In presence of global symmetry breaking, $J_{\mu}^{\text{vis}}\neq0$ in general.
The total U(1) current is conserved:%
\begin{equation}
\partial_{\mu}J_{\mu}=\partial_{\mu}J_{\mu}^{\text{mat}}-v\,\partial^{2}%
\chi=0. \label{total}%
\end{equation}
This is the case if and only if the Lagrangian contains the coupling
\begin{equation}
{\fam\rsfsfam\relax L}\supset\frac{1}{v}\chi\cdot\partial_{\mu}J_{\mu
}^{\text{mat}}. \label{eq:GT}%
\end{equation}
This is sometimes known as Goldberger-Treiman relation, and shows that
Goldstone interactions are predicted in terms of nonconservation of the matter
symmetry current induced in the process of symmetry breaking.

\medskip

When the U(1) symmetry is gauged, the total gauge-invariant Lagrangian is%
\begin{equation}
\Lags=-\frac{1}{4}F_{\mu\nu}^{2}+\frac{1}{2}(\partial_{\mu}\chi-vA_{\mu}%
)^{2}+(A_{\mu}-\frac{1}{v}\partial_{\mu}\chi)J_{\mu}^{\text{mat}%
}+\Lags_{\text{mat}}. \label{eq:Lgauge}%
\end{equation}
We can fix the unitary gauge by setting $\chi$ to zero or, equivalently, by
redefining $A_{\mu}^{\prime}=A_{\mu}-\partial_{\mu}\chi/v$. The second term in
\textrm{(\ref{eq:Lgauge}) }becomes a mass term for $A_{\mu}^{\prime}$. Notice
that while $A_{\mu}$ couples to the total conserved current $J_{\mu}$ given by
eq.~\textrm{(\ref{eq:totalJ})}, the massive vector $A_{\mu}^{\prime}$ couples
to $J_{\mu}^{\text{mat}}$.

\medskip

The last thing that we want to recall concerns production of massive gauge
bosons at high energy $E\gg v$. The effective Lagrangian appropriate for this
situation is obtained from \textrm{(\ref{eq:Lgauge})} by keeping terms with
the highest number of derivatives in $A_{\mu}$ and $\chi$, and is given by
\begin{equation}
\Lags_{\text{HE}}=-\frac{1}{4}F_{\mu\nu}^{2}+\frac{1}{2}(\partial_{\mu}%
\chi)^{2}+A_{\mu}J_{\mu}^{\text{mat}}+\frac{1}{v}\chi\cdot\partial_{\mu}%
J_{\mu}^{\text{mat}}+{\fam\rsfsfam\relax L}_{\text{mat}}.
\end{equation}
We thus see that the total cross section can be approximated (up to terms suppressed by
$v/E$)
by the sum of
massless gauge boson production plus production of Goldstones with coupling
(\ref{eq:GT}):%
\begin{equation}
\sigma(A_{\mu}^{\prime})=\sigma(A_{\mu})+\sigma(\chi). \label{eq:equivGauge}%
\end{equation}
This statement is called the equivalence theorem. It can also be deduced (in a
less transparent way) from the fact that the physical state projector for the
gauge boson of mass $m$ takes the form $-g_{\mu\nu}+k_{\mu}k_{\mu}/m^{2}$.
Notice that, while to get the correct Goldstone production rate it is crucial
to take current nonconservation into account, in computing $\sigma(A_{\mu})$
we can actually assume that $J_{\mu}^{\text{mat}}$ is conserved.
%The error
%induced by neglecting $\partial_{\mu}J_{\mu}^{\text{mat}}\sim v$ in this
%computation is $O(v/E)$ just as the accuracy of the equivalence theorem
%itself. One can estimate this error e.g. by comparing the transverse state and
%the Feynman gauge projectors for the massless gauge boson:
%\[
%\Pi_{\mu\nu}^{\perp}=\varepsilon_{\mu}^{+}\left(  \varepsilon_{\nu}%
%^{+}\right)  ^{\ast}+\varepsilon_{\mu}^{-}\left(  \varepsilon_{\nu}%
%^{-}\right)  ^{\ast}=-\eta_{\mu\nu}-P_{\mu}Q_{\nu}-P_{\nu}Q_{\mu},\quad
%Q_{\mu}=O(1/E).
%\]

The main points of the above discussion --- the form of the total current, the
Goldberger-Treiman relation, the fact that the massive gauge boson couples to
the same matter current, and the equivalence theorem --- will find their
analogues in the gravitino case.

\subsubsection*{Goldstino interaction}

We now repeat the steps in the previous section in the case of supersymmetry,
under which the Goldstino $\chi$ transforms as $\delta\chi=\sqrt
{2}F\varepsilon$, where $F$ is a supersymmetry-breaking vev\footnote{Although
we choose the notations and normalizations which are standard for $F$-term
supersymmetry breaking, the result apply to any combination of $F$ and
$D$-term breaking.} and $\varepsilon$ is the supersymmetric parameter. The
supercurrent is
\begin{equation}
S^{\mu}=S_{\text{vis}}^{\mu}+i\sqrt{2}F\gamma^{\mu}\chi\label{s-full}%
\end{equation}
where the apex `vis' signals that we are interested in theories consisting of
a visible and a hidden sector. Supersymmetry is broken spontaneously in a
heavy hidden sector, and its  low energy remnant is the Goldstino field:
all other hidden sector fields can be integrated out, if one is interested in
energies below the  messenger scale. The full supercurrent is
conserved:
\begin{equation}
\partial_{\mu}S^{\mu}=\partial_{\mu}S_{\text{vis}}^{\mu}+i\sqrt{2}%
F\ds \,\chi=0. \label{eq:dS0}%
\end{equation}
The vanishing of~\textrm{(\ref{eq:dS0})} gives the equation of
motion for $\chi$,
%has to have the
%form%
%\[
%i\partial\hspace{-5pt}{\scriptstyle/}\hspace{1pt}\chi=-\frac{1}{\sqrt{2}%
%F}\partial_{\mu}S_{\text{vis}}^{\mu}%
%\]
and consequently implies the following Goldstino Lagrangian:
\begin{equation}
\Lags_{\text{Goldstino}}=\frac{1}{2}\bar{\chi}i\ds \,\chi
-\frac{1}{\sqrt{2}F}\bar{\chi}\,\partial_{\mu}S_{\text{vis}}^{\mu}%
+\cdots\label{eq:LGoldstino}%
\end{equation}
where $\cdots$ indicates couplings involving two or more Goldstinos, not
needed in our computation.

\subsubsection*{Massless gravitino}

In the supersymmetric limit, the massless gravitino is described by a Majorana
Rarita-Schwinger field $\psi_{\mu}$ with Lagrangian
\begin{equation}
{{\fam\rsfsfam\relax L}}=-\frac{1}{2}\varepsilon^{\mu\nu\rho\sigma}\bar{\psi
}_{\mu}\gamma_{5}\gamma_{\nu}\partial_{\rho}\psi_{\sigma}\label{RS-lagr}%
\end{equation}
invariant under the gauge SUSY transformations with parameter $\varepsilon$:
$\delta\psi_{\mu}=-2\bar{M}_{\text{Pl}}\partial_{\mu}\varepsilon$. Here
$\bar{M}_{\mathrm{Pl}}=M_{\mathrm{Pl}}/8\pi=2.4~10^{18}\,\mathrm{GeV}$ is the
reduced Planck mass. The variation of the matter action defines the Majorana
supercurrent $S_{\mu}$ as
\begin{equation}
\delta S_{\text{matter}}=\int d^{4}x~\bar{S}^{\mu}\partial_{\mu}\varepsilon.
\end{equation}
Demanding that the full action is invariant to zeroth order in $\bar
{M}_{\text{Pl}}^{-1}$ one obtains how the massless gravitino interacts with
the supercurrent
\begin{equation}
{\Lags}_{\text{int}}=\frac{1}{2\bar{M}_{\mathrm{Pl}}}\bar{\psi}_{\mu}S^{\mu
}.\label{grav-int}%
\end{equation}
%no +h.c.

\subsubsection*{Super-Higgs mechanism}

We will now follow how the massless gravitino eats the Goldstino, getting a
mass via the super-Higgs mechanism. First of all, the gauge-invariant action
for the goldstino-gravitino system is~\cite{GraviGold,Cremmer}:
\begin{equation}
\Lags =-\frac{1}{2}\varepsilon^{\mu\nu\rho\sigma}\bar{\psi
}_{\mu}\gamma_{5}\gamma_{\nu}\partial_{\rho}\psi_{\sigma}+\frac{1}{2}\bar
{\chi}i\ds\,\chi-m_{3/2}\bigg[\frac{1}{4}\bar{\psi}_{\mu}[\gamma^{\mu}%
,\gamma^{\nu}]\psi_{\nu}+\bar{\chi}\chi-\sqrt{\frac{3}{2}}\bar{\psi}_{\mu
}i\gamma^{\mu}\chi\bigg].\label{grav-gold}%
\end{equation}
It contains a gravitino-goldstino mixing mass term, that agrees with the form
of the supercurrent, eq.~(\ref{s-full}). Indeed, this Lagrangian is invariant
under the local field transformations:
\begin{align*}
\delta\psi_{\mu} &  =-\bar{M}_{\text{Pl}}(2\partial_{\mu}\varepsilon
+im_{3/2}\gamma^{\mu}\varepsilon),\\
\delta\chi &  =\sqrt{2}F\varepsilon,
\end{align*}
provided that the gravitino mass $m_{3/2}$ and the SUSY breaking vev $F$ are
related as
\begin{equation}
m_{3/2}=\frac{F}{\sqrt{3}\bar{M}_{\text{Pl}}}.\label{eq:Gmass}%
\end{equation}
(The derivation above used the flat space assumption). Introducing the
gravitino mass required a deformation of the supersymmetric transformation of
the gravitino, and the gravitino interaction term with matter (\ref{grav-int})
is no longer invariant. To restore invariance, we must add to the Lagrangian
the term
\begin{equation}
\frac{m_{3/2}}{2\sqrt{2}F}\bar{\chi}i\slashed{S}=\frac{1}{2\bar{M}_{\text{Pl}%
}}\frac{1}{\sqrt{6}}\bar{\chi}i\slashed{S}\qquad\slashed{S}=\gamma_{\mu
}S_{\mathrm{vis}}^{\mu}\label{eq:trace}%
\end{equation}
It may seem surprising at first to find this new coupling of Goldstino to the
supercurrent in addition to the one in (\ref{eq:LGoldstino}). However, there
is no contradiction since the new term vanishes as gravity is decoupled.

We can now choose the unitary gauge $\chi=0$ or equivalently
define\footnote{This notation reflects the fact that $\Psi_{\mu}$ is `bigger'
than $\psi_{\mu}$ since it contains more degrees of freedom.}
\begin{equation}
\Psi_{\mu}=\psi_{\mu}-\frac{1}{\sqrt{6}}i\gamma^{\mu}\chi-\sqrt{\frac{2}{3}%
}\frac{\partial_{\mu}\chi}{m_{3/2}},\label{eq:massive}%
\end{equation}
such that the whole Lagrangian describing gravitino, goldstino, and their
interaction with matter can be rewritten as~\cite{Cremmer}
\begin{equation}
\Lags=-\frac{1}{2}\varepsilon^{\mu\nu\rho\sigma}\bar{\Psi
}_{\mu}\gamma_{5}\gamma_{\nu}\partial_{\rho}\Psi_{\sigma}-\frac{m_{3/2}}%
{4}\bar{\Psi}_{\mu}[\gamma^{\mu},\gamma^{\nu}]\Psi_{\nu}+\frac{1}{2\bar
{M}_{\mathrm{Pl}}}\bar{\Psi}_{\mu}S_{\mathrm{vis}}^{\mu}.\label{eq:full}%
\end{equation}
%This is similar to the standard Higgs mechanism, when the full action could be
%written in terms of $A_{\mu}-\frac{1}{v}\partial_{\mu}\chi$. Alternatively to
%such rewriting, we can simply use local susy invariance to set $\chi=0.$
This is the Lagrangian describing the massive gravitino $\Psi_{\mu}$. We see
that it couples to $S_{\mathrm{vis}}^{\mu}$.

\subsubsection*{Equivalence theorem}

The Lagrangian (\ref{eq:full}) could be used to study production of
massive gravitinos at any energy below the messenger scale. In this
paper, we are interested in energies much bigger than $m_{3/2}$ and
the sparticle masses. A simpler effective Lagrangian appropriate for
these energies can be derived by noticing that the mass terms and
mixings between $\psi_{\mu}$ and $\chi$ in (\ref{grav-gold}) can be
neglected. Thus we can study production of \textit{massless}
gravitinos $\psi_{\mu}$ and Goldstinos $\chi$ coupled to the
visible sector by%
\begin{equation}
\Lags_{\text{int}}=\frac{1}{2\bar{M}_{\text{Pl}}}\bar{S}_{\text{vis}}^{\mu
}(\psi_{\mu}-\sqrt{\frac{1}{6}}i\gamma_{\mu}\chi)+\frac{1}{\sqrt{2}F}\bar
{\chi}\partial_{\mu}S_{\text{vis}}^{\mu}.\label{eq:ETappendix}%
\end{equation}
This is the analogue of the previously mentioned equivalence theorem for
production of gauge bosons at energies much larger than their masses. In the
massive gauge boson case, the equivalence theorem could also be derived from
the form of the physical state projector of the massive gauge boson. Below we
will see that an analogous derivation can also be given for the massive
gravitino case. Just as in the gauge boson case, we can assume that the
supercurrent in the coupling $\bar{S}_{\text{vis}}^{\mu}\psi_{\mu}$ is
conserved; in the Goldstino coupling the current nonconservation is of course
crucial and has to be taken into account.
%We can neglect the gravitino mass and use the polarization tensor... and
%$p\hspace{-4.2pt}{\scriptstyle/}$ for Goldstinos. (In other words, we can
%treat the mass terms as small perturbations). In the MSSM we can drop
%$\slashed{S}$%
%.\textbf{\color{magenta}[\color{red}BUT: CANCELLATION IN \eq{decompo}??
%EFFECTS OF NEW TERM ZERO IN SCATTERINGS AND SUPPRESSED BY $g^2$ at FINITE $T$?\color{magenta}]}
%The supercurrent is given in... and its divergence in.... In practice, this
%means that we should use the projector obtained by keeping only $m_{3/2}^{-2}$
%and $m_{3/2}^{0}$ terms in (\ref{Pi-massive}).

A further simplification concerns the Goldstino coupling (\ref{eq:trace}): as
we explain below this coupling is irrelevant in MSSM at energies much above
the $\mu$-term due to approximate scale-invariance. Thus
\begin{equation}
\Lags_{\text{int}}=\frac{1}{2\bar{M}_{\text{Pl}}}\bar{\psi}_{\mu}%
S_{\text{vis}}^{\mu}+\frac{1}{\sqrt{2}F}\bar{\chi}\partial_{\mu}S_{\text{vis}%
}^{\mu}\qquad \text{ \ (MSSM)}.\label{eq:ET_MSSM}%
\end{equation}

It is instructive to compare the relative importance of the two terms in
(\ref{eq:ET_MSSM}) for the total production rate. Since the divergence of the
supercurrent will be proportional to the soft-breaking masses (see below), the
effective coupling in the second term is $m_{\text{soft}}/F\sim
1/M_{\text{mess}}$. Thus the two terms are equally important if SUSY breaking
is mediated by gravity, $M_{\mathrm{mess}}\sim\bar{M}_{\mathrm{Pl}}$. If
instead $M_{\mathrm{mess}}\ll\bar{M}_{\mathrm{Pl}}$, like in gauge-mediation
models, the Goldstino term dominates~\cite{gravitinoCosmo}.

%\textit{Remark}. In a more complete analysis the action (\ref{full}) comes out
%of the full SUGRA action. The above discussion shows that the relevant terms
%can be derived in a simpler way.

\subsection{The gravitino propagator and polarization tensor}

\subsubsection*{Massless gravitino}

Since the Lagrangian is invariant under local supersymmetry, the same physics
can be described by different choices of gravitino propagators and
polarization tensors. Analogously to the vector case, the sum over the two
physical transverse polarizations is (see \cite{Van})
\begin{equation}
\Pi^{3/2}_{\mu\nu}(P)\equiv\sum_{i=\pm}\Psi_{\mu}^{(i)}\bar{\Psi}_{\nu}^{(i)}%
=\frac{1}{2}\gamma_{\nu}P\hspace{-1.5ex}/\,\gamma_{\mu}
-\frac{1}{2}P_{\mu}\gamma_{\nu}P\hspace{-1.5ex}/\,\slashed{Q}
-\frac{1}{2}P_{\nu}\slashed{Q}P\hspace{-1.5ex}/\gamma^{\mu}
+P_{\mu}P_{\nu}\slashed{Q}
\end{equation}
where $U$ is an arbitrary 4-velocity that defines a preferred reference frame
used to define what `transverse' means, and $Q^{\mu}\equiv\lbrack2U^{\mu
}(U\cdot P)-P^{\mu}]/2(U\cdot P)^{2}$. Gauge-invariant observables do not
depend on the choice of $U$.
%$$
%\frac{1}{2} \left\{\hat{},\gamma _{},\gamma _{}\right\}-\frac{\left\{\hat{},\hat{},\gamma _{}\right\} P_{\mu }}{2 (P\cdot
%V)}-\frac{\left\{\hat{}\right\} P_{\mu } P_{\nu }}{2 (P\cdot V)^2}+\frac{\left\{\hat{},\hat{},\gamma _{}\right\} P_{\nu
%}}{2 (P\cdot V)}+\frac{\left\{\hat{}\right\} P_{\mu } V_{\nu }}{(P\cdot V)}-\left\{\hat{}\right\} \eta _{\mu \nu }$$

\smallskip

As usual, local gauge invariance allows to define more convenient gauge
choices. For example, one can impose the gauge-fixing condition $F\equiv
\gamma^{\mu}\Psi_{\mu}=0$ (in this gauge one also has $\partial_{\mu}\Psi
^{\mu}=0$ as a consequence of the equations of motion). To derive the
propagator, it is best to consider an analogue of the $\xi$-gauge by adding to
the Lagrangian the gauge-fixing term $\bar{F} i \ds \,F/\xi$. Then the kinetic
operator is invertible, and the gravitino propagator is~\cite{Das}
\begin{equation}
\frac{\Pi^{3/2}_{\mu\nu}}{P^{2}+i\varepsilon}\qquad\hbox{with}\qquad\Pi^{3/2}_{\mu\nu
}=\frac{1}{2}\gamma^{\nu}P\hspace{-1.5ex}/\,\gamma^{\mu}-(2+\xi)\frac{P^{\mu
}P\hspace{-1.5ex}/\,\,P^{\nu}}{P^{2}}. \label{ProjN}%
\end{equation}
As usual, $\Pi^{3/2}_{\mu\nu}$ is also the projector to be used when the massless
gravitino production rate is summed over the gravitino polarizations. The
dependence on the gauge-fixing parameter $\xi$ is irrelevant because the
massless gravitino couples to the conserved supercurrent, $P_{\mu}S_{\mu}=0$.
For the simplest choice $\xi=-2$ one has
\begin{equation}
\Pi^{3/2}_{\mu\nu}=\frac{1}{2}\gamma_{\nu}P\hspace{-1.5ex}/\,\gamma_{\mu}=-\frac
{1}{2}\gamma_{\mu}P\hspace{-1.5ex}/\,\gamma_{\nu}-P\hspace{-1.5ex}/\,\eta
_{\mu\nu}+\gamma_{\mu}P_{\nu}+P_{\mu}\gamma_{\nu}. \label{eq:PiN}%
\end{equation}
The last two terms do not contribute, again because the supercurrent is
conserved. For our later computation we will choose
\begin{equation}\label{eq:PiSenzaP}
\Pi^{3/2}_{\mu\nu}=-\frac{1}{2}\gamma_{\mu}P\hspace{-1.5ex}/\,\gamma_{\nu}%
-P\hspace{-1.5ex}/\,\eta_{\mu\nu}.
\end{equation}

\subsubsection*{Massive gravitino and the equivalence theorem}

The massive gravitino is described by the Lagrangian (\ref{eq:full}). The mass
term breaks the gauge symmetry present in the massless case. The equations of
motion coming from the free part of~\textrm{(\ref{eq:full})} imply
\begin{equation}
\gamma_{\mu}\Psi_{\mu}=0,\qquad\partial_{\mu}\Psi_{\mu}=0,\qquad
(\slashed{P}-m_{3/2})\Psi_{\mu}=0.\label{on-shell}%
\end{equation}
In the massless case the first two equations could have been imposed as
gauge-fixing conditions. The resulting propagator is~\cite{Van} $\Pi_{\mu\nu
}/(P^{2}-m_{3/2}^{2}+i\varepsilon)$ where
\begin{equation}
\Pi_{\mu\nu}=-(\slashed{P}+m_{3/2})\left(  g_{\mu\nu}-\frac{P_{\mu}P_{\nu}%
}{m_{3/2}^{2}}\right)  -\frac{1}{3}\left(  \gamma^{\mu}+\frac{P^{\mu}}%
{m_{3/2}}\right)  (\slashed{P}-m_{3/2})\left(  \gamma^{\nu}+\frac{P^{\nu}%
}{m_{3/2}}\right)  .\label{eq:Pimassive}%
\end{equation}
Again $\Pi_{\mu\nu}$ is also the polarization tensor to be used when summing
over all physical polarizations: $\Pi_{\mu\nu}=\sum_{i=\pm\frac{1}{2},\pm
\frac{3}{2}}\Psi_{\mu}^{(i)}\bar{\Psi}_{\nu}^{(i)}$. One can check
that~\textrm{(\ref{eq:Pimassive})} is consistent with (\ref{on-shell}).

In this paper we are interested in production of ultrarelativistic gravitinos.
To study this limit, we expand~\textrm{(\ref{eq:Pimassive}) }in powers of
$m_{\text{3/2}}$:
\begin{align}
\Pi_{\mu\nu}= &  ~\frac{2}{3}\frac{P_{\mu}P_{\nu}\slashed{P}}{m_{3/2}^{2}%
}+\frac{4P_{\mu}P_{\nu}-\slashed{P}\gamma^{\nu}P^{\mu}-\gamma^{\mu
}\slashed{P}P^{\nu}}{3m_{3/2}}\nonumber\\
&  ~+\left(  -g_{\mu\nu}\slashed{P}-\frac{1}{3}\gamma_{\mu}\slashed{P}\gamma
_{\nu}+\frac{1}{3}\gamma^{\mu}P^{\nu}+\frac{1}{3}\gamma^{\nu}P^{\mu}\right)
+\left(  \frac{\gamma^{\mu}\gamma^{\nu}}{3}-g^{\mu\nu}\right)  m_{3/2} .
\label{Pi-massive}%
\end{align}
If the supercurrent to which the gravitino couples is conserved, the terms
singular in $m_{3/2}$ give no contribution. The last term vanishes for
$m_{3/2}\rightarrow0$. The term that does not depend on $m_{3/2}$
\emph{differs} from the massless gravitino projector (\ref{eq:PiN}) by
$\gamma_{\mu}\slashed{P}\gamma_{\nu}/6$, modulo irrelevant terms proportional
to $P_{\mu}$ or $P_{\nu}$.

Thus in the limit $m_{3/2}\rightarrow0$ we not only recover the massless
gravitino, but also get an additional massless spin $1/2$ fermion which
couples to ${1}/\sqrt{6}$ times the `trace' of the supercurrent,
$\slashed{S}=\gamma_{\mu}S^{\mu}$. This is akin to the van
Dam-Veltman-Zakharov discontinuity~\cite{vVZ} encountered when adding to the
graviton a Fierz-Pauli mass term $m_{g}$: the limit $m_{g}\rightarrow0$ then
describes the usual massless graviton plus a scalar coupled to the trace of
the energy-momentum tensor $T_{~\mu}^{\mu}$. In our case this `discontinuity'
is entirely expected and is consistent with the equivalence theorem as
expressed by eq.~(\ref{eq:ETappendix}): the extra spin 1/2 fermion is nothing but the Goldstino.

\medskip

When we take soft SUSY-breaking into account, the supercurrent is no
longer conserved.
The first term in (\ref{Pi-massive}) can then be interpreted as
corresponding to the Goldstino production due to the last term in
(\ref{eq:ETappendix}) (the coefficient agrees as one checks using
(\ref{eq:Gmass})). In this derivation of the equivalence theorem it
is non-obvious that the terms in (\ref{eq:Pimassive}) proportional
to $m_{3/2}^{-1}$ should cancel, as is required for full agreement
with~(\ref{eq:ETappendix}). However this cancellation does
happen, as verified in the explicit computations needed for this paper.

\subsection{MSSM supercurrent at zero temperature}

\subsubsection*{Gravitino couplings}

In a generic renormalizable SUSY gauge theory with vector supermultiplets
$(A_{\mu}^{a},\lambda^{a})$ and matter chiral supermultiplets $\Phi_{i}%
=(\phi_{i},\xi_{i})$ and superpotential $W$ the Weyl part $s^{\mu}$ of the
Majorana supercurrent $S^{\mu}=(s^{\mu}$,$\bar{s}^{\mu})$ is (see
e.g.\ \cite{Drees} or explicitly compute it)
\begin{equation}
s^{\mu}=-\sqrt{2}\left[  (D^{\nu}\phi_{i})^{\ast}(\sigma_{\nu}\bar{\sigma
}^{\mu}\xi_{i})+iW_{i}^{\ast}(\phi^{\ast})\sigma^{\mu}\bar{\xi}_{i}\right]
-\frac{1}{2}F_{\nu\rho}^{a}(\sigma^{\nu}\bar{\sigma}^{\rho}\sigma^{\mu}%
\bar{\lambda}^{a})-ig(\phi_{i}^{\ast}T_{ij}^{a}\phi_{j})(\sigma^{\mu}%
\bar{\lambda}^{a}) \label{eq:2-comp}%
\end{equation}
where $D_{ij}^{\mu}=\delta_{ij}\partial_{\mu}+igA^{\mu a}T_{ij}^{a}$ is the
gauge-covariant derivative. The first two terms are the supercurrent of the
Wess-Zumino model and of the SUSY gauge theory without matter. The third term
is a correction which arises as a result of coupling between the two. (With
the Noether formalism it would arise because the Lagrangian is supersymmetric
up to a total derivative). In 4-component notation it becomes\footnote{See
\cite{Weinberg}, page 141. Notice that~\cite{Weinberg} has a misprint in
normalizing the second line of the RHS of (27.4.40), cf (26.7.10). The
difference in $\gamma^{5}$ in the terms involving gluinos is because our
gluino-squark-quark coupling is real: $\lambda_{\text{our}}=i\gamma^{5}%
\lambda_{\text{his}}$. The extra $i$ is then compensated by the difference in
$\gamma_{\text{our}}^{\mu}=i\gamma_{\text{his}}^{\mu}$.}:%
\begin{align}
S^{\mu}=  &  -\sqrt{2}\left[  (D^{\nu}\phi_{i})^{\ast}(\gamma^{\nu}\gamma
^{\mu}\xi_{L}^{i})+(D^{\nu}\phi_{i})(\gamma^{\nu}\gamma^{\mu}\xi_{R}%
^{i})-iW_{i}(\phi)\gamma^{\mu}\xi_{L}^{i}-iW_{i}^{\ast}(\phi^{\ast}%
)\gamma^{\mu}\xi_{R}^{i}\right] \label{eq:4-comp}\\
&  -\frac{1}{4}F_{\nu\rho}^{a}[\gamma^{\nu},\gamma^{\rho}]\gamma^{\mu}%
\gamma^{5}\lambda^{a}-ig(\phi_{i}^{\ast}T_{ij}^{a}\phi_{j})\gamma^{\mu}%
\lambda^{a}\nonumber
\end{align}
where we introduced Majorana spinors ($\xi_{i},\bar{\xi}_{i})^{T}$ and
$(\lambda^{a},\bar{\lambda}^{a})^{T}$ which by abuse of notation we denoted
again $\xi_{i}$ and $\lambda^{a}$. As usual $\xi_{L}\equiv P_{L}\xi$, $\xi
_{R}\equiv P_{R}\xi$, $W_{i}=\partial W/\partial\phi_{i}$ and the index $i$
runs over all chiral multiplets.

The supercurrent is conserved $\partial_{\mu}S^{\mu}=0$ as a
consequence of equations of motion. After fixing the vector gauge
symmetries in the usual way, the vector equations of motion change
due to the gauge-fixing terms and to the ghost current. The ghosts
are scalars under supersymmetry (in particular, they do not have
superpartners and they couple only to the gauge field but not to the
gaugino), and one could be worried that SUSY is broken by the gauge
choice. Indeed the supercurrent divergence is no longer zero,
however it is BRST exact (see e.g.~\cite{BRST}). Thus the amplitude
for longitudinal gravitino emission still vanishes, and the
gravitino gauge invariance is preserved.
%The last terms of each line do not
%contribute to massive gravitino production because $\gamma^{\mu}\Psi_{\mu}=0$
%on shell, but must be taken into account when computing production of massless
%gravitinos using the previously discussed gauge fixing.

\smallskip

The terms proportional to $\gamma^{\mu}$ are sometimes omitted from the
supercurrent expression (\ref{eq:4-comp}), because they do not contribute to
the massive gravitino production due to the on-shell condition $\gamma^{\mu
}\Psi_{\mu}=0$. However, one should be careful to keep these terms if one
wants to use the equivalence theorem and the massless gravitino gauge
invariance, because the supercurrent is no longer conserved if they are omitted.

\subsubsection*{Goldstino couplings}

According to the equivalence theorem discussed above, the spin $\pm1/2$
component of the massive gravitino at high energies can be replaced by the
Goldstino coupled to the divergence and trace $\slashed{S}$ of the visible
sector supercurrent with coefficients given in (\ref{eq:ETappendix}).
%. We have%
%\begin{equation}
%S_{\text{goldstino}}=\frac{1}{\sqrt{2}F}\bar{\chi}\cdot\partial_{\mu}S^{\mu}.
%\end{equation}
The divergence $\partial_{\mu}S_{\text{vis}}^{\mu}$ measures the SUSY breaking
in the visible sector, which at energies lower than the messenger scale looks
like explicit breaking by soft terms. In absence of soft terms $\partial_{\mu
}S_{\text{vis}}^{\mu}=0$ as a consequence of equations of motion. Nonzero soft
terms modify the equations of motion, so that $\partial_{\mu}S_{\text{vis}%
}^{\mu}\sim m_{\text{soft}}\neq0$. For dimensional reasons we can neglect
dimension 2 soft terms (i.e.\ scalar squared masses): only soft terms with
dimension 1 (i.e.\ gaugino masses $M$ and trilinear scalar couplings $A$)
contribute to Goldstino production at dominant order, $\gamma\propto T^{6}$.
By taking into account how the relevant soft terms modify the equations of
motion of particles and sparticles we get
%To obtain the goldstino coupling, we have to compute $\partial_{\mu}S^{\mu}$
%taking into account that the equations of motion of gauginos and scalars get
%modified in presence of the soft-breaking terms by the following correction
%terms:%
%\begin{equation}
%\Delta(i\partial\!\!\!\raisebox{2pt}[0pt][0pt]{$\scriptstyle/$}\,\lambda
%^{a})=m_{\lambda}\lambda^{a},\qquad\Delta(\partial^{2}\tilde{q})=-\lambda
%_{t}A_{t}(h\tilde{t})^{\ast}\text{ }\&~\text{permutations.}%
%\label{eq:gluino-mass}%
%\end{equation}
%We can treat the gauge and $\lambda_{t}$ effects independently. For the gauge
%term we find, differentiating (\ref{eq:s_qcd}) and taking
%eq.~(\ref{eq:gluino-mass}) into account:%
\begin{equation}
\partial_{\mu}S_{\text{vis}}^{\mu}=-\frac{iM}{4}F_{\nu\rho}^{a}[\gamma^{\nu
},\gamma^{\rho}]\gamma^{5}\lambda^{a}-Mg(\phi^{\ast}T^{a}\phi)\lambda
^{a}+\sqrt{2}[(AW)_{i}(\phi)\xi_{L}^{i}+(AW)_{i}^{\ast}(\phi^{\ast})\xi
_{R}^{i}] \label{eq:dS}%
\end{equation}
where $i$ runs over all chiral multiplets and a sum is understood over the
components of the gauge group. The presence of the second gauge term was first
noticed in~\cite{LeeWu}, and is here reobtained via a simple direct
computation. Notice that to get it, it is crucial to keep the last term in
eq.~\textrm{(\ref{eq:4-comp})}, that does not contribute to massive gravitino
production due to the on-shell condition $\gamma^{\mu}\Psi_{\mu}=0$.

In the MSSM the relevant soft terms are the three gaugino masses
$M_{1,2,3}$ and the top $A$-term, $A_{t}$:
\begin{equation}
\Lags_{\text{soft}}=\sum_{N=1}^{3} \frac{M_{N}}{2} \lambda^{a}_{N}\lambda
^{a}_{N}+ \lambda_{t}A_{t}(\tilde{Q}\tilde{U} H_{\mathrm{u}}%
+\hbox{h.c.})+\cdots.
\end{equation}

\bigskip

%For the superpotential coupling (\ref{eq:W_top}) we find, differentiating
%(\ref{eq:s_WZ}):
%\[
%\partial_{\mu}S^{\mu}=\sqrt{2}\lambda_{t}A_{\text{t}}[W_{i}(\phi)\xi_{L}%
%^{i}+W_{i}^{\ast}(\phi^{\ast})\xi_{R}^{i}]\text{ \ \ (}\lambda_{t}\text{
%contribution)}%
%\]
Finally, we elaborate on the Goldstino coupling to $\slashed{S}$, finding that
it can be neglected in the MSSM. Using $\gamma^{\mu}[\gamma^{\nu},\gamma
^{\rho}]\gamma^{\mu}=0$, (\ref{eq:2-comp}) implies%
\begin{equation}
\bar{\sigma}^{\mu}s^{\mu}=-\sqrt{2}\left[  -2(D^{\nu}\phi_{i})^{\ast}%
\sigma_{\nu}\xi_{i}+4iW_{i}^{\ast}(\phi^{\ast})\bar{\xi}_{i}\right]
-4ig(\phi_{i}^{\ast}T_{ij}^{a}\phi_{j})\bar{\lambda}^{a}%
\end{equation}
Rewriting the first term as
\[
(D^{\nu}\phi_{i})^{\ast}\sigma_{\nu}\xi_{i}=\partial_{\nu}(\phi_{i}^{\ast
}\sigma_{\nu}\xi_{i})-\phi_{i}^{\ast}\sigma_{\nu}D_{\nu}\xi_{i}%
\]
and using the fermion equation of motion, several terms cancel and we remain
with%
\begin{equation}
\bar{\sigma}^{\mu}s^{\mu}=2\sqrt{2}\partial_{\nu}(\phi_{i}^{\ast}\sigma_{\nu
}\xi_{i})+2\sqrt{2}i[\phi_{i}^{\ast}W_{ij}^{\ast}-2W_{j}^{\ast}(\phi^{\ast
})]\bar{\xi}_{j}\label{Sslash}%
\end{equation}
The first term does not contribute to massless Goldstino production
rate since
$\bar{\chi}\partial_{\nu}(\phi_{i}^{\ast}\sigma_{\nu}\xi_{i})$
vanishes on-shell due to $\bar{\sigma}_{\nu}\partial_{\nu}\chi=0.$
The second term vanishes if $W_{j}$ is a quadratic function of the
fields, i.e.\ for cubic terms in $W$. We thus conclude that the only
nontrivial coupling to Goldstino arising from the (\ref{eq:trace})
vertex is due to the
$\mu$-term and is of the form%
\[
\sim\frac{\mu}{\bar{M}_{\text{Pl}}}\chi(H_{1}\tilde{H}_{2}+\tilde{H}_{1}%
H_{2})+\text{h.c.}%
\]
This vertex is irrelevant at energies much bigger than $\mu$.

The reason for the above is that the trace of the supercurent $\slashed{S}$
falls into a supersymmetric ``anomaly multiplet"
\[
\{\slashed{S},\partial_{\mu}R^{\mu},T_{\,\mu}^{\mu}\},
\]
where $T_{\,\mu}^{\mu}$ is the trace of the energy-momentum tensor expressing
the scale invariance of the theory, and $R^{\mu}$ is the current of the
$R$-symmetry under which all chiral multiplets have charge $2/3$ (see
\cite{Weinberg}). In MSSM, both the scale invariance and the above
$R$-symmetry are broken classically only by the $\mu$-term, and this explains
$\slashed{S}\sim\mu$. At quantum level the scale invariance and the
$R$-symmetry are anomalous, e.g. $\partial_{\mu}R^{\mu}$ is given by the
triangle anomaly equation:%
\[
\partial_{\mu}R^{\mu}=\sum\frac{b_{N}g_{N}^{2}}{48\pi^{2}}F_{\mu\nu}%
^{(N)}\tilde{F}_{\mu\nu}^{(N)}%
\]
where the anomaly coefficients $b_{N}=\{11,1,-3\}$ are the same as the
one-loop $\beta$-function coefficients of the MSSM gauge groups, which is
again related to the fact that $\partial_{\mu}R^{\mu}$ and $T_{\,\mu}^{\mu}$
are in the same supermultiplet. Since supersymmetry relates $\slashed{S}$ to
$\partial_{\mu}R^{\mu}$, one can show that (see \cite{West})
\begin{equation}
\slashed{S}=\sum\frac{b_{N}g_{N}^{2}}{16\pi^{2}}F_{\mu\nu}^{(N)}[\gamma^{\mu
},\gamma^{\nu}]\lambda^{(N)}.\label{anomaly}%
\end{equation}
Below we  argue that this equation can be used also at finite temperature.

\subsection{Gravitino and goldstino couplings at finite temperature}
Gravitino production from a supersymmetric thermal plasma is best
studied in terms of its non-time ordered propagator $\Pi^{<}(P)$ given by eq.~(\ref{eq:Pi<}). Supersymmetry is broken by finite temperature, but this
breaking is spontaneous, so the supercurrent remains conserved:
$\partial_{\mu}{}S_{\mu}=0$ holds as an operator equation. This means that the
production rate of longitudinal gravitinos vanishes also at finite
temperature. Equivalently, $\Pi^{<}(P)$ is invariant under gauge
transformations of the gravitino polarization tensor,
$
\delta\Pi_{\mu\nu}^{3/2}=P_{\mu}A_{\nu}+P_{\nu}B_{\mu}$,
with arbitrary $A,B$. The statements of the previous paragraph should hold identically in any
computation including all diagrams to a given order in the thermal bath
coupling $g$. In practice, however, it may be difficult to see the vanishing
of $\delta\Pi^{<}$ explicitly. E.g.\ as explained in section~\ref{subtractions} we
are resumming a well-defined class of physical effects to order $g^{4}$:
those enhanced by a $1\to 2$ phase space factor, unlike a generic
${\cal O}(g^{4})$ correction. More precisely, diagram D is computed including thermal
corrections to the propagators of particles to which the gravitino couples,
while diagrams S$_{1,2,3}$ are computed using tree-level propagators.
In particular, we do not include corrections to the gravitino vertex.
For this reason we expect a residual gravitino gauge dependence, which we believe
to be of relative order $g^{2}/\pi^2$ with respect to our result. The reason is that
in our calculation of the massless gravitino production rate, thermal masses act
similar to soft SUSY-breaking terms, modifying equations of motion by terms of
order $g^{2}T$, so that $\partial\cdot S\sim g^{2}T$ rather than being zero.
This means that $\delta\Pi^{<}\sim {\cal O}(g^{4})$.
We see that this non-gauge invariance is of the right order of
magnitude to be cancelled by vertex corrections.
The above residual non-gauge invariance can be tolerated when computing
the massless gravitino production rate.

When computing the Goldstino
production rate, we have taken into account that,
in absence of soft-SUSY breaking,
 the Goldstino coupling to $\partial\cdot S$ vanishes at finite temperature
 by evaluating  the divergence of the supercurrent before
computing the thermal matrix element, i.e.\ we start the
finite-temperature computation from eq. (\ref{eq:dS}). Since we do
not evaluate vertex corrections, this procedure is expected to give
a result with the same ${\cal O}(g^{2}/\pi^2)$ error as the
gravitino production rate.

Finally, the anomaly relation (\ref{anomaly}) valid at zero
temperature also holds at finite temperature.
The argument is the same as in case of the supercurrent conservation:
the thermal bath is a background, and~(\ref{anomaly})
is a dynamical property of the Hamiltonian valid for any background. In
practice this means that the Goldstino coupling to $\slashed{S}$ can be neglected.

\setcounter{equation}{0}
\section{Vector propagator at finite temperature}\label{thermal}
We list the full one-loop expressions for thermal corrections to a vector
with four-momentum $K=(\omega,\vec{k})$ ($K^2=\omega^2-k^2$)
with respect to the  rest frame of the thermal plasma.
In general, we denote by $U_\mu$ the four-velocity $U_\mu$ of the plasma.
We use the Feynman gauge where all effects are condensed in two
form factors even in the non-abelian case~\cite{WeldonVector}.
Polarizations are conveniently decomposed in
 $T$ransverse
(i.e.\ orthogonal to $K$ and to $\vec{k}$), $L$ongitudinal
(i.e.\ orthogonal to $K$ and parallel to $\vec{k}$) and
parallel to $K$.
The corresponding projectors $(\Pi^T +\Pi^L +\Pi^K)_{\mu \nu}=-\eta_{\mu\nu}$ are
\begin{eqnsystem}{sys:PiV}
\Pi_{\mu\nu}^T &=&-\tilde{\eta}_{\mu\nu}+\frac{\tilde{K}_\mu\tilde{K}_\nu}{-k^2}=
 \begin{pmatrix}
0 & 0\cr 0 & \delta_{ij}- k_i k_j/k^2 \end{pmatrix},\\
 \Pi_{\mu\nu}^L &=& -\eta_{\mu\nu} + \frac{K_\mu K_\nu}{K^2} -\Pi_{\mu\nu}^T ,\\
\Pi_{\mu\nu}^K  &=&  - \frac{K_\mu K_\nu}{K^2},
\end{eqnsystem}
where $\tilde{\eta}_{\mu\nu}=\eta_{\mu\nu}-U_\mu U_\nu$, $\tilde{K}_{\mu} = K_\mu - (K\cdot U) U_\mu$.
The vector propagator is
\beq^*D_{\mu\nu} =  i\bigg[\frac{\Pi_{\mu\nu}^T}{K^2-\pi_0- \pi_T} + \frac{\Pi_{\mu\nu}^L}{K^2-\pi_0-\pi_L} +  \frac{\Pi_{\mu\nu}^K}{K^2}\bigg].\eeq
In the following $\simeq$ denotes the HTL limit,
where the result can be expressed in terms of the  vector thermal mass $m_V^2 = \frac{1}{6}g^2 T^2 (N+N_S + N_F/2)$, where the $V$ector, $F$ermion and $S$calar coefficients  are defined having in mind a group $\SU(N)$ with $N_F$ massless
Dirac fermions and $N_S$ scalars plus anti-scalars  in the fundamental representation. Table~\ref{tab:NNN} lists the explicit values of $N,N_F,N_S$ in the SM and in the MSSM.
The one-loop quantum correction at $T=0$ in the $\overline{\rm MS}$ scheme is
% COMPUTATION OF THE SCALAR COEFFICIENT.
% the ratios for fermions and scalars are the same as in gauge beta functions.
% in the SM H contributes to the beta function 1/2  than L.
% we are using units where both L and H count as "N=1/2"
%%\footnote{\color{red}
%%ADD $N_S$, FIX FACTORS USING [WELDON]
%%$$
%%\Pi_{\mu\nu} = g^2 \frac{-5N+2 N_F}{48\pi^2} \ln \frac{-K^2}{\bar\mu^2}
%%(K^2 \eta_{\mu\nu}-K_\mu K_\nu)$$}
\begin{eqnsystem}{sys:vector}
\pi_0 &=&  g^2 K^2 \frac{2 N_F+N_S-5N}{48\pi^2} \ln \frac{-K^2}{\bar\mu^2}   \label{eq:pi0}\\
\riga{where the gauge-dependent vector loop gives a negative contribution to spectral densities
above the light cone, at $K^2>0$.
%{\em\color{red} MATTER WINS, but $\pi_0$ is 10 times smaller than $\pi_{L,T}$
%above the light cone: is the $48\pi^2$ factor correct?.}
The thermal corrections are}\\
\pi_L &=&  -\frac{K^2}{k^2} g^2 (N_S H_S + N_F H_F + N H_V) \simeq
-\frac{K^2}{k^2} (L+1) m_V^2 ,\\
\pi_T &=& -\frac{\pi_L}{2} + \frac{g^2}{2}(N_S G_S + N_F G_F + N G_V)\simeq m_V^2 (1+\frac{K^2}{k^2}\frac{L+1}{2})\\
\riga{where}\\[-2ex]
G_S &=& \int_0^\infty \frac{dp}{2\pi^2}\bigg[4p - \frac{K^2}{4k} L_- \bigg] n_B(p)\simeq \frac{T^2}{3} \\
G_F &=& \int_0^\infty \frac{dp}{2\pi^2}\bigg[4p + \frac{K^2}{2k}L_- \bigg] n_F(p) \simeq \frac{T^2}{6}\\
G_V &=& \int_0^\infty \frac{dp}{2\pi^2}\bigg[4p+\frac{5K^2}{4k}L_- \bigg]n_B(p) \simeq \frac{1}{3}T^2\\
H_S &=& \int_0^\infty \frac{dp}{2\pi^2}\bigg[2p L
+\frac{M}{k}  +\frac{k}{4} L_- \bigg]n_B(p) \simeq \frac{L+1}{6}T^2  \\
H_F &=& \int_0^\infty \frac{dp}{2\pi^2}\bigg[2pL+\frac{M}{k}
 \bigg] n_F (p)\simeq  \frac{L+1}{12}T^2\\
H_V &=& \int_0^\infty \frac{dp}{2\pi^2}\bigg[ 2pL+\frac{M}{k} -\frac{k}{4} L_- \bigg]n_B(p) \simeq \frac{L+1}{6}T^2
\end{eqnsystem}
having defined $\omega_\pm\equiv (\omega\pm k)/2$,
$$L \equiv  1-\frac{\omega}{k}\ln\frac{\omega_+}{\omega_-},\qquad
L_\pm  \equiv  \ln\frac{p+\omega_+}{p+\omega_-}\pm\ln\frac{p-\omega_+}{p-\omega_-},$$
$$M  \equiv  (p+\omega_+)(p+\omega_-)\ln\frac{p+\omega_+}{p+\omega_-}-(p-\omega_+)(p-\omega_-)\ln\frac{p-\omega_+}{p-\omega_-}\simeq 2kp$$
See~\cite{WeldonVector,thoma} for previous results. We added scalar
loops and wrote logarithms such that imaginary parts (needed to get
spectral densities) are obtained using the prescription $\omega\to
\omega + i0^+$, with $\ln z$ having a cut along the negative real
axis. We emphasize that our expressions cannot be simplified using
$\ln(a b) = \ln a + \ln b$, because this would give wrong imaginary
parts. The spectral densities employed in eq.\eq{rhoV} are defined
as \beq \rho_T = -2\Im  \frac{1}{K^2 -\pi_0-\pi_T},\qquad \rho_L
=-2\Im \frac{K^2}{k^2}\frac{1}{K^2-\pi_0-\pi_L}. \eeq As well
known~\cite{WeldonVector,LeBellac} $\rho_L$ contains a collective
longitudinal excitation, that corresponds to longitudinal waves of
electric fields allowed by Maxwell equations with vanishing
dielectric constant.

\begin{table}[t]
\begin{center}
$$\begin{array}{c|ccc|ccc}
& \multicolumn{3}{|c|}{\hbox{Standard Model}} & \multicolumn{3}{|c}{\hbox{MSSM}}\\
\hbox{Vector} & N & N_F & N_S & N & N_F & N_S\\ \hline
\hbox{Gluon $\SU(3)_c$} & 3 &6 &0& 3&9&6\\
\hbox{Weak $\SU(2)_L$}&2 & 6 &1/2 & 2 & 9 &7\\
\hbox{Hypercharge ${\rm U}(1)_Y$} &0 & 10 & 1/2 & 0 & 11 & 11
\end{array}$$
\end{center}
\caption{\label{tab:NNN}\em Numerical coefficients for vector thermal mass $m_V^2 = \frac{1}{6}g^2 T^2 (N+N_S + N_F/2)$.}
\end{table}%

\setcounter{equation}{0}
\section{Fermion propagator at finite temperature}\label{thermalF}
Fermions can receive thermal corrections from gauge and Yukawa couplings.
In HTL approximation the full result is determined, in a generic non-supersymmetric theory,
by one parameter, the thermal mass:
\beq
m_F^2 = \bigg[ \frac{C_R}{8} g^2 + \frac{\lambda^2}{16}\bigg]T^2\eeq
where we used for $g,\lambda,C_R$ the same notation as in\eq{MTSUSY},
except that $g,\lambda$ here denote non-supersymmetric couplings.
The same parameter $m_F$ controls the full one-loop expression in the Feynman gauge.
See~\cite{WeldonFermion,thoma} for previous results.
%Real parts have been computed in~\cite{WeldonFermion}.
%Our results are basically identical, except that our expressions
%can also be used to compute imaginary parts.
The spectral densities for particles ($\rho_+$) and holes ($\rho_-$) are given by
%\beq
%\rho_+ = -2 \Im\bigg[\omega - k - \frac{m_F^2}{k}(K+U(k-\omega))\bigg]^{-1},\qquad
%% \frac{1}{A_0 - A_S},\qquad
%\rho_- = - 2\Im\bigg[\omega + k + \frac{m_F^2}{k}(K-U(k+\omega))\bigg]^{-1}
%% \frac{1}{A_0 + A_S}
%\eeq
%%where
%%$$A_0 = \omega - m_F^2 U,\qquad A_S = k + \frac{m_F^2}{k} (K - \omega U)$$
%where
%\begin{eqnsystem}{sys:fermion}
%K(\omega,k) &=& \int_0^\infty \frac{dp}{\pi^2}  \frac{\omega^2-k^2}{2k} L_- [n_F(p)-n_B(p)] ,\\
%U(\omega,k) &=&  \int_0^\infty \frac{dp}{\pi^2} \frac{1}{k}\bigg[-L_+ p[n_F(p) + n_B(p)] - L_- \omega n_B(p)\bigg] .
%\end{eqnsystem}
\beq \rho_\pm = - \Im\bigg[\omega_\mp \bigg(1 - \frac{1}{2\pi^2}
(\frac{C_R}{8} g^2+ \frac{\lambda^2}{16}) \ln \frac{-K^2}{\bar\mu^2}\bigg)
+ m_F^2 F_\pm
  \bigg]^{-1}\eeq
where the $T=0$ contribution gives a spectral density only above the light-cone, and
\beq
F_\pm (\omega,k) =
\mp  \int_0^\infty \frac{dp}{\pi^2}\frac{\omega_\mp}{k^2}
\bigg[ pL_+\cdot  (n_B(p)+n_F(p)) + L_- \cdot (n_B(p)\omega_- +n_F(p)\omega_+ )\bigg]
%  \int_0^\infty \frac{dp}{\pi^2}\frac{\omega_\mp}{k^2}
%\bigg[ k L_- [n_B(p)-n_F(p)] \mp (2 p L_+ + \omega L_-)[n_B(p)+n_F(p)]\bigg]
\mp \frac{L \omega_\mp + \omega_\pm}{k\omega}.
\eeq
The functions $\omega_\pm = (\omega \pm k)/2$
and $L_\pm$ are the same previously defined for vectors.
The last terms is the HTL contribution (complex only below the light-cone).
Again branch cuts are defined by the prescription $\omega\to \omega + i 0^+$.

\footnotesize

\bigskip\bigskip

\begin{multicols}{2}
\end{multicols}

\begin{thebibliography}{nn}

\bibitem{gravitinoCosmo}
\art{J.R.~Ellis, J.~E.~Kim, D.V.~Nanopoulos}{\PL}{B145}{181}{1984}.
\art[hep-ph/9403364]{M. Kawasaki, T. Moroi}{Prog. Theor. Phys.}{93}{879}{1995}.
See also \hepart[hep-ph/9503210]{T. Moroi}.
\art[hep-ph/9809381]{M. Bolz, W. Buchmuller, M. Plumacher}{Phys. Lett.}{B443}{209}{1998}.


\bibitem{Buch}
\art[hep-ph/0012052]{M. Bolz, A. Brandenburg, W. Buchmuller}{Nucl. Phys.}{B606}{518}{2001}.


\bibitem{postBuch}
\hepart[hep-ph/0608344]{J. Pradler, F.D. Steffen}.
\hepart[hep-ph/0612291]{J. Pradler, F.D. Steffen}.


\bibitem{BY} \art{E. Braaten, T.C. Yuan}{\PRL}{66}{2183}{1991}.


\bibitem{LeBellac}
M. Le Bellac, {\em Thermal Field Theory}, Cambridge University Press (2000).
%, ISBN0521654777.


\bibitem{Pisarski92}
\art{E. Braaten, R.D. Pisarski}{\PR}{D45}{R1827}{1992}.


\bibitem{CE}
\art[hep-ph/9606438]{D. Comelli, J. Espinosa}{Phys. Rev.}{D55}{6253}{1997}.


\bibitem{WeldonFermion}
Fermions at finite temperature.
V.V. Klimov, Sov. J. Nucl. Phys. 33, 934 (1981);
H.~A.~Weldon, Phys.\ Rev.\ D {26}, 2789 (1982)
and Phys.\ Rev.\ D {40}, 2410 (1989).


\bibitem{WeldonVector}
{Vectors at finite temperature}.
D.~J.~Gross, R.~D.~Pisarski and L.~G.~Yaffe,
%``QCD And Instantons At Finite Temperature,''
Rev.\ Mod.\ Phys.\ {53}, 43 (1981).
\art{H. Weldon}{\PR}{D26}{1394}{1982}.
H.~T.~Elze, K.~Kajantie and T.~Toimela,
%``CHROMOMAGNETIC SCREENING AT HIGH TEMPERATURE,''
Z.\ Phys.\ C {37}, 601 (1988).
R.~Kobes, G.~Kunstatter and K.~W.~Mak,
%``LINEAR RESPONSE OF THE HOT QCD PLASMA FROM THE GLUON PROPAGATOR,''
Z.\ Phys.\ C {45}, 129 (1989).
\art[hep-ph/9701279]{H. Weldon}{Annals Phys.}{271}{141}{1999}.
%\art[hep-ph/9908204]{H. Weldon}{Phys. Rev.}{D61}{036003}{2000}.


\bibitem{thoma}
\art[hep-ph/9708434]{A. Peshier, K. Schertler, M. Thoma}{Annals Phys.}{266}{162}{1998}.


\bibitem{wrong}
  \art[hep-th/9404044]{W. Fischler}{Phys. Lett.}{B332}{277}{1994}.


\bibitem{LR}
  \art[hep-ph/9503402]{R. Leigh, R. Rattazzi}{Phys. Lett.}{B352}{20}{1995} and
  \art[hep-ph/9505438]{J. Ellis, D. Nanopoulos, K. Olive, S. Rey}{Astropart. Phys.}{4}{371}{1996}
showed that thermal effects do not give contributions to the gravitino production rate
of the form $\gamma\approx T^8/m_{3/2}^2M_{\rm Pl}^2$.
The first paper also showed that the mixing of the true Goldstino
with the thermal Goldstino can be ignored.


\bibitem{Ellis}
\art[hep-ph/9505438]{J. Ellis, D. Nanopoulos, K. Olive, S. Rey}{Astropart. Phys.}{4}{371}{1996}.


\bibitem{SUSYT}
Supersymmetry at finite temperature.
 \art{D. Boyanovsky}{\PR}{D29}{743}{1984}.
 \art{H. Aoyama}{\PL}{B171}{420}{1986}.
 \art{R. Gudmundsdottir, P. Salomonson}{\NP}{B285}{1}{1987}.
\art[hep-th/0303260]{K. Kratzert}{Ann. Phys.}{308}{285}{2003}.
See also among the references of these papers.
%Other incorrect or out of focus works about the issue of  supersymmetry at finite temperature
%can be found among the references of these works.
We are aware of no works containing explicit results for the supercurrent at finite temperature.

\bibitem{largeNf}
See \art[hep-ph/0301057]{A. Ipp, G. Moore, A. Rebhan}{JHEP}{01}{037}{2003}
for a recent discussion and references.


\bibitem{Starinets}
  S.~Caron-Huot, P.~Kovtun, G.~D.~Moore, A.~Starinets and L.~G.~Yaffe,
  %``Photon and dilepton production in supersymmetric Yang-Mills plasma,''
  JHEP {0612}, 015 (2006)
  [hep-th/0607237].
  
  
\bibitem{flat}
%\art[hep-ph/0512227]{R. Allahverdi, A. Mazumdar}{JCAP}{0610}{008}{2006}.
For recent discussions see
\art[hep-ph/0608096]{K. Olive, M. Peloso}{Phys. Rev.}{D74}{103514}{2006} and
\hepart[hep-ph/0608296]{R. Allahverdi, A. Mazumdar}.



\bibitem{book} See e.g.\
E. W. Kolb, M. S. Turner, {\it The Early
Universe},
(Addison-Wesley, Menlo Park, Ca., 1990).


\bibitem{leptog}
The constraint on the reheating temperature for successful thermal MSSM leptogenesis was found to be
$T_{\rm RH}> 1.6~10^{9}\GeV$ in
\art[hep-ph/0310123]{G.F. Giudice, A. Notari, M. Raidal, A. Riotto, A. Strumia}{Nucl. Phys.}{B685}{89}{2004},
working in one flavor approximation and warning that this approximation is generically accurate up to ${\cal O}(1)$ corrections.
The constraint was reconsidered in  \hepart[hep-ph/0611232]{S. Antusch, A.M. Texeira},
where flavor was included (and some ${\cal O}(g^2/\pi^2)$ corrections neglected),
finding  $T_{\rm RH}>1.9~10^{9}\GeV$.


\bibitem{WMAP}
\hepart[astro-ph/0603449]{D.N.~Spergel {\it et al.} (WMAP Science Team)}.


\bibitem{Moroi}
\art[hep-ph/0507245]{K. Kohri, T. Moroi, A. Yotsuyanagi}{Phys. Rev.}{D73}{123511}{2006}.
\hepart[hep-ph/0605215]{M. Pospelov}.


\bibitem{Viel}
\art[astro-ph/0605706]{M. Viel, J. Lesgourgues, M. Haehnelt, S. Matarrese, A. Riotto}{Phys. Rev. Lett.}{97}{071301}{2006}.


\bibitem{altri}
For recent works, see e.g.\
\art[hep-ph/0512044]{K. Jedamzik, K. Choi, L. Roszkowski, R. Ruiz de Austri}{JCAP}{0607}{007}{2006}.
\art[hep-ph/0607261]{J. Ellis, A. Raklev, O. Oye}{JHEP}{10}{061}{2006}.
\hepart[astro-ph/0608562]{R.~H.~Cyburt, J.~Ellis, B.~D.~Fields, K.~A.~Olive and V.~C.~Spanos}{JCAP}{0611}{014}{2006}.
%\art[hep-ph/0312262]{J.R.~Ellis, K.~A.~Olive, Y.~Santoso, V.~C.~Spanos}{\PL}{B588}{7}{2004}.


\bibitem{DI}
We here discuss the leptogenesis constraint on the reheating temperature.
It is implied by  the Davidson-Ibarra bound
\art[hep-ph/0202239]{S. Davidson, A. Ibarra}{Phys. Lett.}{B535}{25}{2002},
that only holds up to an ${\cal O}(1)$ flavor factor
(see e.g.\
\art[hep-ph/0601084]{E. Nardi, Y. Nir, E. Roulet, J. Racker}{JHEP}{01}{164}{2006} and
\hepart[hep-ph/0605281]{A. Abada, S. Davidson, F. Josse-Michaux, M. Losada, A. Riotto} for a recent discussion)
as already claimed in
\art[hep-ph/9911315]{R. Barbieri, P. Creminelli, N. Tetradis, A. Strumia}{\NP}{B575}{61}{2000}.
Furthermore it holds assuming that right-handed neutrinos are very hierarchical:
thermal leptogenesis at low temperature is possible within the standard see-saw
if right-handed neutrinos  are mildly hierarchical (see e.g.\
\art[hep-ph/0408015]{M. Raidal, A. Strumia, K. Turzynski}{Phys. Lett.}{B609}{351}{2005})
or quasi-degenerate
(see e.g.\
\art{M. Flanz, E.A. Paschos, U. Sarkar and J. Weiss}{\PL}{B389}{693}{1996};
\art{L. Covi, E. Roulet}{\PL}{B399}{113}{1997};
\art[hep-ph/0309342]{A. Pilaftsis, T. Underwood}{Nucl. Phys.}{B692}{303}{2004}).


\bibitem{Lykken}
\hepart[hep-th/9612114]{J.~D.~Lykken}.
  %``Introduction to supersymmetry,''



\bibitem{GraviGold}
S.~Deser and B.~Zumino,
  %``Broken Supersymmetry And Supergravity,''
  Phys.\ Rev.\ Lett.\  {38}, 1433 (1977).


\bibitem{Cremmer} E.~Cremmer, B.~Julia, J.~Scherk, S.~Ferrara, L.~Girardello
and P.~van Nieuwenhuizen,
  %``Spontaneous Symmetry Breaking And Higgs Effect In Supergravity Without
  %Cosmological Constant,''
  Nucl.\ Phys.\ B {147}, 105 (1979).


\bibitem{Van}
 P.~Van Nieuwenhuizen,
  %``Supergravity,''
  Phys.\ Rept.\  {68}, 189 (1981).


\bibitem{Das}
 A.~Das and D.~Z.~Freedman,
  %``Gauge Quantization For Spin 3/2 Fields,''
  Nucl.\ Phys.\ B {114}, 271 (1976).


\bibitem{vVZ}
 H.~van Dam and M.~J.~G.~Veltman,
  %``Massive And Massless Yang-Mills And Gravitational Fields,''
  Nucl.\ Phys.\ B {22}, 397 (1970).
  V.~I.~Zakharov,
  %``Linearized gravitation theory and the graviton mass,''
  JETP Lett.\  {12}, 312 (1970)
  [Pisma Zh.\ Eksp.\ Teor.\ Fiz.\  {12}, 447 (1970)].


\bibitem{Drees}
  M.~Drees, R.~Godbole and P.~Roy,
  ``{\em Theory and phenomenology of sparticles: An account of four-dimensional $N=1$
  supersymmetry in high energy physics}'', World Scientific, 2004.
%\href{http://www.slac.stanford.edu/spires/find/hep/www?irn=6240364}{SPIRES entry}


\bibitem{Weinberg}
  S.~Weinberg,
  ``{\it The quantum theory of fields.  Vol. 3: Supersymmetry}''.
%\href{http://www.slac.stanford.edu/spires/find/hep/www?irn=4384008}{SPIRES entry}


\bibitem{BRST}
\art[hep-th/9708007]{K.~Fujikawa and K.~Okuyama}{\NP}{B521}{401}{1998}.
  %``BRST gauge fixing and the algebra of global supersymmetry,'' .


\bibitem{LeeWu}
%\art{P. Fayet}{\PL}{84B}{421}{1979}.
\art{T. Lee, G.-H. Wu}{\PL}{B447}{83}{1999}.


\bibitem{West}
  P.~C.~West, ``{\em Introduction to supersymmetry and supergravity}'', World Scientific, 1990.


\end{thebibliography}
\end{document}